\documentclass[acmsmall,screen]{acmart}

\usepackage{amsfonts}
\usepackage{algorithmic}
\usepackage{algorithm}
\usepackage{balance}
\usepackage[table]{colortbl}
\usepackage{longtable}
\usepackage{makecell}
\usepackage{multirow}
\usepackage{pdfsync}
\usepackage{stfloats}
\usepackage[caption=false,font=footnotesize]{subfig}
\usepackage{tabularray}
\usepackage{tabularx}
\usepackage{textcomp}
\usepackage{threeparttable}
\usepackage{url}
\usepackage{verbatim}
\usepackage{xcolor}

\AtBeginDocument{%
  }

\begin{document}

\title{Resource Allocation and Workload Scheduling for Large-Scale Distributed Deep Learning: A Survey}

\author{Feng Liang}
\email{fliang@smbu.edu.cn}
\orcid{0000-0002-8542-9871}
\affiliation{%
  \institution{Artificial Intelligence Research Institute, Shenzhen MSU-BIT University}
  \country{China}
  \postcode{518107}
}
\affiliation{%
  \institution{Guangdong-Hong Kong-Macao Joint Laboratory for Emotional Intelligence and Pervasive Computing, Shenzhen MSU-BIT University}
  \country{China}
  \postcode{518107}
}

\author{Zhen Zhang}
\email{zhangzhen19@lzu.edu.cn}
\orcid{0009-0007-9955-0916}
\author{Haifeng Lu}
\email{luhf18@lzu.edu.cn}
\orcid{0000-0003-0155-8447}
\affiliation{%
  \institution{Gansu Provincial Key Laboratory of Wearable Computing, School of Information Science and Engineering, Lanzhou University}
  \country{China}
  \postcode{730000}
}
\affiliation{%
  \institution{Guangdong-Hong Kong-Macao Joint Laboratory for Emotional Intelligence and Pervasive Computing, Shenzhen MSU-BIT University}
  \country{China}
  \postcode{518107}
}

\author{Chengming Li}
\email{licm@smbu.edu.cn}
\orcid{0000-0002-8542-9871}
\affiliation{%
  \institution{Artificial Intelligence Research Institute, Shenzhen MSU-BIT University}
  \country{China}
  \postcode{518107}
}
\affiliation{%
  \institution{Guangdong-Hong Kong-Macao Joint Laboratory for Emotional Intelligence and Pervasive Computing, Shenzhen MSU-BIT University}
  \country{China}
  \postcode{518107}
}

\author{Victor C. M. Leung}
\email{vleung@ieee.org}
\orcid{0000-0003-3529-2640}
\affiliation{%
  \institution{Artificial Intelligence Research Institute, Shenzhen MSU-BIT University}
  \country{China}
  \postcode{518107}
}
\affiliation{%
  \institution{Department of Electrical and Computer Engineering, The University of British Columbia}
  \country{Canada}
  \postcode{V6T 1Z4}
}

\author{Yanyi Guo}
\authornote{Corresponding authors.}
\email{guoyy@smbu.edu.cn}
\orcid{0009-0000-7682-6667}
\affiliation{%
  \institution{Frontier Cross Disciplinary Research Institute, Shenzhen MSU-BIT University}
  \country{China}}
\affiliation{%
  \institution{School of Mechanical and Electrical Engineering, Beijing Institute of Technology}
  \country{China}}
  \postcode{10081}

\author{Xiping Hu}
\authornotemark[1]
\email{huxp@smbu.edu.cn}
\orcid{0000-0002-4952-699X}
\affiliation{%
  \institution{Artificial Intelligence Research Institute, Shenzhen MSU-BIT University}
  \country{China}
  \postcode{518107}
}
\affiliation{%
  \institution{Guangdong-Hong Kong-Macao Joint Laboratory for Emotional Intelligence and Pervasive Computing, Shenzhen MSU-BIT University}
  \country{China}
  \postcode{518107}
}
\affiliation{%
  \institution{School of Medical Technology, Beijing Institute of Technology}
  \country{China}
  \postcode{10081}
}

\renewcommand{\shortauthors}{Liang et al.}

\begin{abstract}
With rapidly increasing distributed deep learning workloads in large-scale data centers, 
efficient distributed deep learning framework strategies for resource allocation and workload scheduling have become the key to high-performance deep learning.
The large-scale environment with large volumes of datasets, models, and computational and communication resources raises various unique challenges for resource allocation and workload scheduling in distributed deep learning, such as scheduling complexity, resource and workload heterogeneity, and fault tolerance. 
To uncover these challenges and corresponding solutions, this survey reviews the literature, mainly from 2019 to 2024, on efficient resource allocation and workload scheduling strategies for large-scale distributed DL.
We explore these strategies by focusing on various resource types, scheduling granularity levels, and performance goals during distributed training and inference processes. 
We highlight critical challenges for each topic and discuss key insights of existing technologies.
To illustrate practical large-scale resource allocation and workload scheduling in real distributed deep learning scenarios, we use a case study of training large language models. 
This survey aims to encourage computer science, artificial intelligence, and communications researchers to understand recent advances and explore future research directions for efficient framework strategies for large-scale distributed deep learning. 

\end{abstract}




\maketitle

\section{Introduction}

With the rapid increase in the sizes of datasets and deep learning (DL) models, distributed DL~\cite{liang2024communicationefficient,ddlCommAlgoSurvey23} has become the state-of-the-art practice for various artificial intelligence technologies, federated learning~\cite{dmlToFL22} and smart Internet of Things~\cite{dlIoTSecuSurvey21}. 
In contrast to traditional single-node DL that works on a single computing node or even a single GPU, distributed DL can leverage multiple GPUs and computing nodes to handle massive training and inference workloads and improve learning throughput. 
Notably, in the era of extremely large models with tens of billions of parameters, distributed DL enables efficient large-model training~\cite{paLM23} across hundreds of computing nodes with thousands of GPUs in the data center.

However, distributed DL faces numerous critical challenges related to efficient framework strategies for resource allocation and workload scheduling in large-scale environments. 
Firstly, with a large number of computational and communication devices in the data center for distributed DL, managing and allocating resources efficiently for distributed DL workloads to utilize resources fully becomes challenging. 
This challenge is amplified in heterogeneous resource environments, where GPUs have various computational capacities and networks have various communication capacities and topologies.
Secondly, distributed DL workloads exhibits more complicated characteristics than those of single-node DL. 
On the one hand, various parallelism modes of distributed DL workloads give rise to new communication patterns involving significant communication overhead for data transfer and model synchronization.
On the other hand, the combination of many computational and communication tasks in distributed DL workloads complicates the execution dependency paradigm, which allows for significant optimization.
Thirdly, the exponential increase in large model sizes raises concerns about the cost of computational and communication resources and efficiency of distributed training in a large scale.
Tackling these challenges is urgent and requires researchers in the fields of computer sciences, artificial intelligence, and communications to understand critical problems in this domain systematically. 

Several existing surveys~\cite{EdegAI19,scalableDL20,commEdgeAI20,commOptDDL21,communicationFL21,ddlCommAlgoSurvey23,commDL23,commDDL23,dlWorkloadScheduling24} have touched on some topics of efficient resource allocation and workload scheduling strategies for distributed DL. 
For example, Ye~\textit{et al.} have introduced scheduling distributed training and inference workloads on GPUs at the job level. 
However, these surveys lack a systematic exploration of distributed DL framework strategies for scheduling computational and communication resources and workloads at various granularity levels in large-scale environments. 
Researchers in the fields of computer science, artificial intelligence, and communications need a comprehensive understanding of representative and critical challenges for framework strategies in large-scale distributed DL environments.  

To fill the gap in existing surveys on distributed DL framework strategies, this survey systematically investigates critical challenges and efficient distributed DL strategies for resource allocation and workload scheduling.
We review the literature over the period mainly between the year 2019 and 2024.
The discussion covers various resource types, scheduling granularity levels, and performance goals. 
For resource allocation strategies, we discuss GPU sharing technologies applying different approaches and network bandwidth-sharing technologies working at different granularity levels. 
For workload scheduling strategies, we categorize the technologies based on various performance goals and scheduling granularities.
Both sets of strategies organized primarily based on the application stage: distributed training and inference.
Focusing on efficient strategies for large-scale distributed DL, we highlight key challenges for each topic and provide insights about crucial research outputs. 
To illustrate how to apply these efficient framework strategies practically in real life, we conduct a case study on large-model distributed training, a rapidly trending and probably long-lasting research topic and one of the best application scenarios that require efficient distributed DL framework strategies. 
We also provide outlooks for future research directions.

The major contribution of this survey is summarized as follows.  
\begin{itemize}
        \item We thoroughly and comprehensively survey up-to-date resource allocation and workload scheduling framework strategies for large-scale distributed DL.
        \item We highlight critical challenges for each topic of these framework strategies.
        \item We use a case study on large-model training to illustrate how to apply efficient framework strategies in practice. 
\end{itemize}



\newcolumntype{Y}{>{\centering\arraybackslash}X}
\newcommand\tabTitle[1]{\cellcolor{green!15}\textbf{#1}}
\begin{table*}[!t]
\tiny
    \caption{A Comparison of Related Surveys}\label{tab:survey}
    \centering
    \begin{tabularx}{\textwidth}{ cc| *{2}{Y}|*{3}{Y} }
    \Xhline{2\arrayrulewidth}
         \tabTitle{}& \tabTitle{} & \multicolumn{2}{c|}{\cellcolor{green!30}\textbf{Resource Allocation}} & \multicolumn{3}{c}{\cellcolor{green!30}\textbf{Workload Scheduling}}  \\
        \tabTitle{Ref.} & \tabTitle{Year} & \tabTitle{GPU Allocation} &\tabTitle{Network Bandwidth Allocation} & \tabTitle{Job Scheduling} & \tabTitle{Pipeline Scheduling} & \tabTitle{Network Flow Scheduling} \\
        \Xhline{2\arrayrulewidth}
        \cite{scalableDL20} & 2020 &  \checkmark & &\checkmark& &\\ 
        \rowcolor{green!5}
        \cite{commEdgeAI20} & 2020 & &\checkmark&&&\\ 
        \rowcolor{green!5} 
        \cite{commOptDDL21} & 2021 &  &&&\checkmark&\checkmark\\ 
        \cite{ddlCommAlgoSurvey23} & 2023 &  &&&\checkmark&\checkmark\\ 
        \rowcolor{green!5}
        \cite{commDL23} & 2023 & &\checkmark&&& \\ 
        \cite{commDDL23} & 2023 & &&&\checkmark& \\ 
        \rowcolor{green!5}
        \cite{dlWorkloadScheduling24} & 2024 & \checkmark &&\checkmark&& \\ 
        \cite{liang2024communicationefficient} & 2024 &&\checkmark&&&\checkmark\\
    \Xhline{2\arrayrulewidth}
        Ours & - & \checkmark & \checkmark & \checkmark & \checkmark & \checkmark \\ 
    \Xhline{2\arrayrulewidth}
    \end{tabularx}
\end{table*}

\subsection{Related Surveys}
Existing surveys on distributed DL lack systematic coverage over strategies for resource allocation of various resource types and workload scheduling at various levels. 
Table~\ref{tab:survey} compares our survey with other related surveys on the covered topics in the domain of distributed DL framework strategies, including resource allocation based on the GPU and network bandwidth and workload scheduling based on the job, pipeline, and network flow. 

Some surveys focus on scheduling distributed jobs on GPU data centers. 
Both Mayer and Jacobsen~\cite{scalableDL20} and Ye~\textit{et al.}~\cite{dlWorkloadScheduling24} have conducted surveys on job-level GPU allocation and workload scheduling for distributed training and inference in the data center environment. 
However, these surveys concern only about job-level strategies focusing on the overall performance of the entire data center but not finer-grained strategies focusing on individual job performance. 
They also only investigate technologies related to the single resource type, the GPU, but not the network bandwidth, which is a significant performance factor for distributed DL with communication as the bottleneck. 

In contrast to the GPU-centric surveys, some works focus on communications technologies and network bandwidth-allocation strategies for distributed DL. 
Both Shi~\textit{et al.}~\cite{commEdgeAI20} and Cao~\textit{et al.}~\cite{commDL23} have reviewed the literature on bandwidth allocation strategies for federating learning over wireless networks.
Liang~\textit{et al.}~\cite{liang2024communicationefficient} have not only investigated bandwidth-allocation strategies on general networks but also studied network-flow-scheduling strategies on different network layers. 
However, covering communications-only technologies does not reveal the whole picture of efficient scheduling in distributed DL, which requires the joint optimization of computation and communication.

For finer-grained workload scheduling in distributed DL, some works~\cite{commOptDDL21,ddlCommAlgoSurvey23,commDDL23} explore pipeline-level scheduling strategies for overlapping computation and communication workloads to improve throughput. 
However, they do not highlight primary challenges related to this topic and lack the investigation into job-level resource allocation and workload scheduling strategies, which can orchestrate with these pipeline-level strategies as a synthesis framework solution for distributed DL in data centers.

Our survey fills the gap in existing surveys. 
We instigate resource allocation strategies for both computational and communication resources, primarily the GPU and network bandwidth, to match the resource requirements of distributed DL workloads. 
We also explore workload scheduling strategies at the job, pipeline, and network flow levels to improve both overall data center throughput and individual job efficiency.
The systematic literature study involving various resource types and scheduling granularity levels makes this article a comprehensive survey of up-to-date technologies in the distributed DL framework domain.

\begin{figure*}[!t]
\centering
\includegraphics[width=1\textwidth]{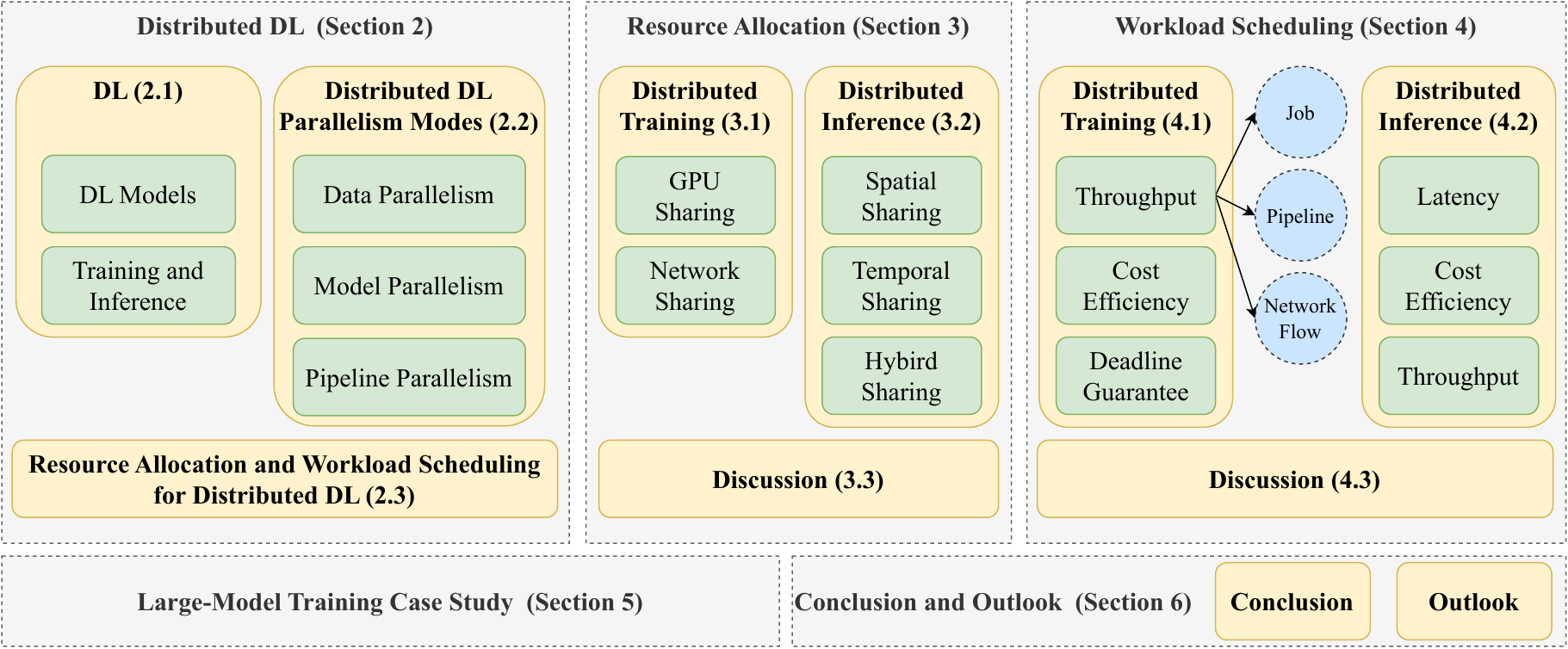}
\caption{The organization of the survey}
\label{fig:organization}
\end{figure*}

\subsection{Survey Organization}
Fig.~\ref{fig:organization} outlines the organization of the remaining sections in this survey. 
Section~\ref{sec:background} provides fundamental knowledge about distributed DL and the resource-management and workload scheduling framework. 
Sections~\ref{sec:resourceAlloc} and~\ref{sec:taskSchedule} introduce various framework strategies for resource allocation and workload scheduling, respectively.
These framework strategies are categorized primarily based on their application scenarios, including distributed training and inference, and secondarily based on resource types, approaches, or performance goals.
We discuss the insights at the end of each section. 
We use a case study of distributed large-model training to show how to apply these framework strategies practically in real-life data centers in Section~\ref{sec:caseStudy}.
In Section~\ref{sec:conclusion}, we conclude this survey and present outlooks of future research directions. 

\begin{table}[!t]
\tiny
\caption{A List of Frequently Used Abbreviations}\label{tab:abbr}
\centering
\begin{tabular}{cl}
\Xhline{2\arrayrulewidth}
\cellcolor{green!15}\textbf{Abbreviation} & \cellcolor{green!15}\textbf{Description}\\
\Xhline{2\arrayrulewidth}
DL & Deep Learning\\
DRL & Deep Reinforcement Learning\\
DNN & Deep Neural Network\\
GPU & Graphics Processing Unit\\
LLM & Large Language Model\\
MPS & Multiple Process Sharing\\
NLP & Natural Language Processing\\
PS & Parameter Server\\
SLO & Service-Level Objective\\
\Xhline{2\arrayrulewidth}
\end{tabular}
\end{table}

\section{Fundamentals of DL and Distributed DL}\label{sec:background}
In this section, we introduce the fundamental knowledge of DL, distributed DL, and the resource allocation and workload scheduling framework for distributed DL.
Table~\ref{tab:abbr} lists frequently used abbreviations used in this survey.

\subsection{Deep Learning}
DL is a subfield of machine learning that utilizes deep artificial neural networks, also known as deep neural networks (DNN), to extract complex patterns from training data in a hierarchical manner. 
The trained DNN is capable to recognize/predict patterns in unseen data. 
DL has been used in various fields, including NLP~\cite{transformerTranslate15}, computer vision~\cite{imageRecog20}, and biomedical engineering~\cite{deepSkinLesion23}.

\begin{figure*}[!t]
\centering
\subfloat[Fully Connected DNN]{\includegraphics[width=0.27\textwidth]{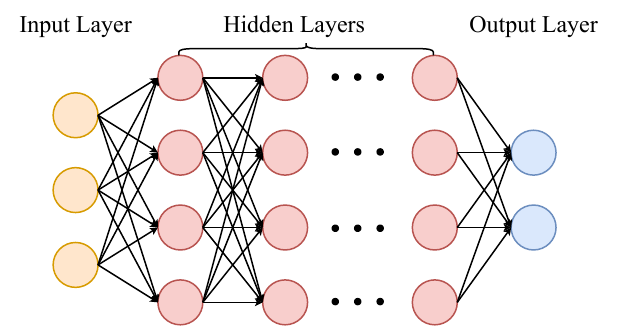}%
\label{fig:model_dnn}}
\hfil
\subfloat[CNN]{\includegraphics[width=0.3\textwidth]{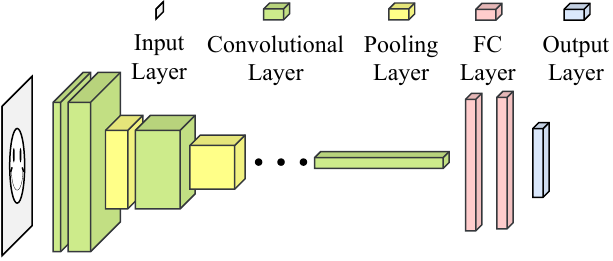}%
\label{fig:model_cnn}}
\hfil
\subfloat[RNN]{\includegraphics[width=0.4\textwidth]{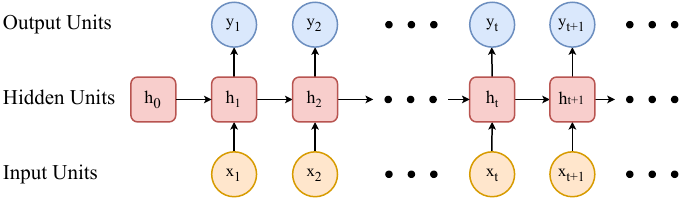}%
\label{fig:model_rnn}}
\caption{Common artificial neural network models for DL}
\label{fig:model}
\end{figure*}

\subsubsection{DL models}
A DNN consists of multiple hidden layers. Each layer is comprised of neurons, which are typically activated by non-linear functions. 
Based on the connections between neurons within and between layers, there can be various types of DNN models.
In this survey, when referring to models or DL models, we mean DNNs unless the context otherwise specifies.
Fig.~\ref{fig:model} illustrates three basic DNN models: the fully connected DNN, convolutional neural network (CNN), and recurrent neural network (RNN). 

$\bullet$ \textbf{Fully connected DNN:} The fully connected DNN, also known as the feedforward neural network, constitutes a dense network with an input layer, a number of hidden layers, and an output layer, as depicted in Fig.~\ref{fig:model_dnn}. 
Neurons in a preceding layer connect to all neurons in the subsequent layer, and each connection has a learnable weight parameter indicating the strength of the connection. 
This architecture enables the fully connected DNN to capture complex relationships within data, finding extensive application in tasks such as classification~\cite{commDnnClassification23}, regression~\cite{commDnnRegression22}, and feature representation embedding~\cite{commDnnFeature20}. 

$\bullet$ \textbf{CNN:} CNN stands as a prevalent model designed for feature extraction and classification, primarily tailored for image and video data. 
As depicted in Fig.~\ref{fig:model_cnn}, 
CNN comprises a stack of convolutional layers and pooling layers for context feature extraction.
Unlike the fully connected layer, which assigns a weight parameter to each neuron connection, the convolutional layer substantially reduces the number of weight parameters by utilizing a number of kernels, each containing shared weights for feature extraction. 
The feature extraction process of the convolutional layer's feature extraction process is empowered by the convolution operation, wherein kernels traverse the receptive fields of an image, extracting new features through weighted summations followed by a non-linear activation function. 
The pooling layer
downsamples the data in the convolutional layer to reduce feature dimensions and alleviate overfitting issues. 
CNN has found widespread applications in various computer vision tasks, including image classification~\cite{snsCnnImage19}, semantic segmentation~\cite{transformerCnnRemoteSensing22}, and object detection~\cite{vehicleCnnObject22}. 

$\bullet$ \textbf{RNN:} RNN, a DL model that deals with sequential data like time-series data, natural language, and speech audio, is illustrated in Fig.~\ref{fig:model_rnn}.
The general architecture of RNN 
includes hidden units that capture and propagate temporal context from the input sequence to subsequent hidden unites. 
It updates continuously and utilizes the temporal context based on the current input and previous temporal context to make predictions.
To address the challenge of capturing long-range temporal dependencies, 
two common variants of RNNs, known as Long Short-Term Memory (LSTM) and Gated Recurrent Unit (GRU), have been developed,
providing a trade-off between modeling such dependencies and reducing computation complexity effectively.
Common applications of RNN include tasks such as time series forecasting~\cite{timeSeriesRnn19}, NLP~\cite{secureNlp20}, and automated planning~\cite{automatedPlanRnn21}.

\subsubsection{Training and inference}
The training of a DL model is the process of optimizing its parameters to minimize the prediction error on a training dataset, as determined by a specified loss function, or objective function. 
Loss functions can be either convex or non-convex, leading to convex or non-convex optimization problems.
Training can be decomposed into two key processes: feedforward and backpropagation. 
In the feedforward process, training data are passed into the model's input layer, and the output prediction is computed by forwarding data through the network using the current model parameters. 
In the backpropagation process, the prediction error and gradients are calculated with respect to the loss function, and trainable parameters are updated iteratively in a backward manner, optimizing the model for the minimum loss.
Common optimizers for backpropagation updating include mini-batch Stochastic Gradient Descent (SGD)~\cite{minibatchSGD14}, SGD with momentum~\cite{momentumSgd13}, Adagrad~\cite{adagrad11}, and Adam~\cite{adam15}.
The training process usually operates on batches of training data iteratively over multiple epochs until the model converges.
A model is said to have converged when the training error settles to within a predefined error range, and additional training will not further decrease the error. 
After completing the training process, the weight parameters in the DL model are learned and fixed. 
Following the training process, there is typically a validation process for validating the performance of the trained model, providing information for fine-tuning hyperparameters and retraining the model for better performance.

The inference process passes forward unseen data through the trained DL model to make predictions. 
Depending on the specific requirements of an application, the resulting prediction can be extracted either either from the output layer or from the predicted latent representation in an intermediate hidden layer. 
For example, in the context of network traffic analysis, an end-to-end DNN model may be trained to classify traffic types directly~\cite{trafficClassificaiton20}, 
Alternatively, an encoder-decoder model trained on traffic data can utilize the latent representation generated by the encoder for subsequent tasks such as attack detection~\cite{latentRepresentationIot20}. 

Computing tasks related to a specific portion of the DL model for specific epochs during the training or inference process are generally referred to as DL tasks in this survey, when it is not necessary to distinguish training tasks and inference tasks in the context.  

\subsection{Distributed DL Parallelism Modes}
\begin{figure*}[!t]
\centering
\subfloat[Data Parallelism]{\includegraphics[width=0.3\textwidth]{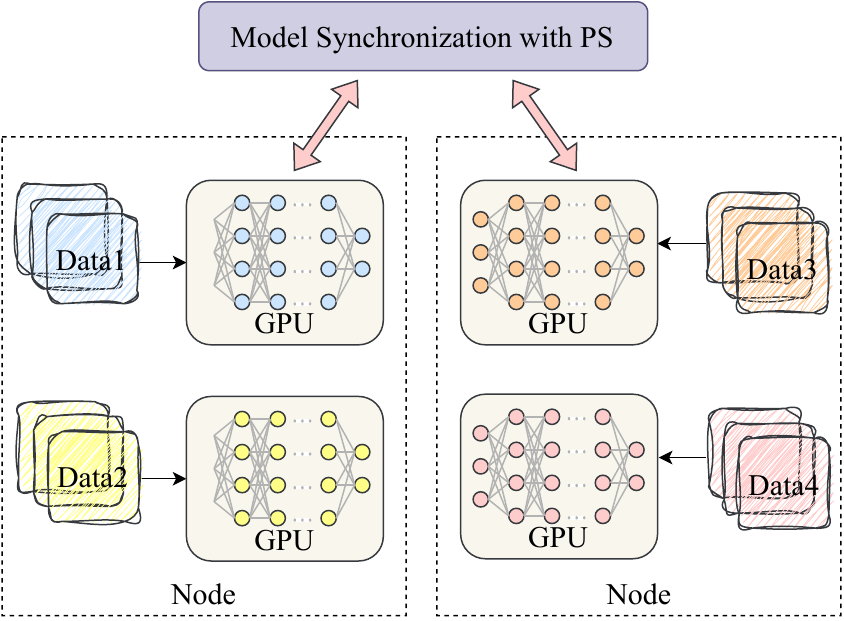}%
\label{fig:parallelism_data}}
\hfil
\subfloat[Model Parallelism]{\includegraphics[width=0.26\textwidth]{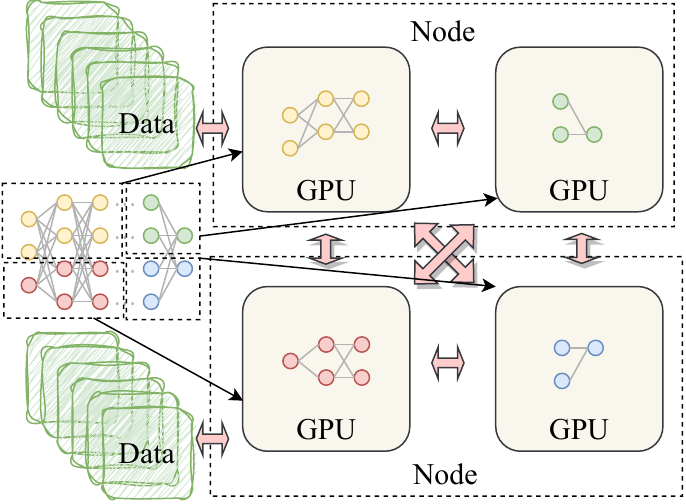}%
\label{fig:parallelism_model}}
\hfil
\subfloat[Pipeline Parallelism]{\includegraphics[width=0.36\textwidth]{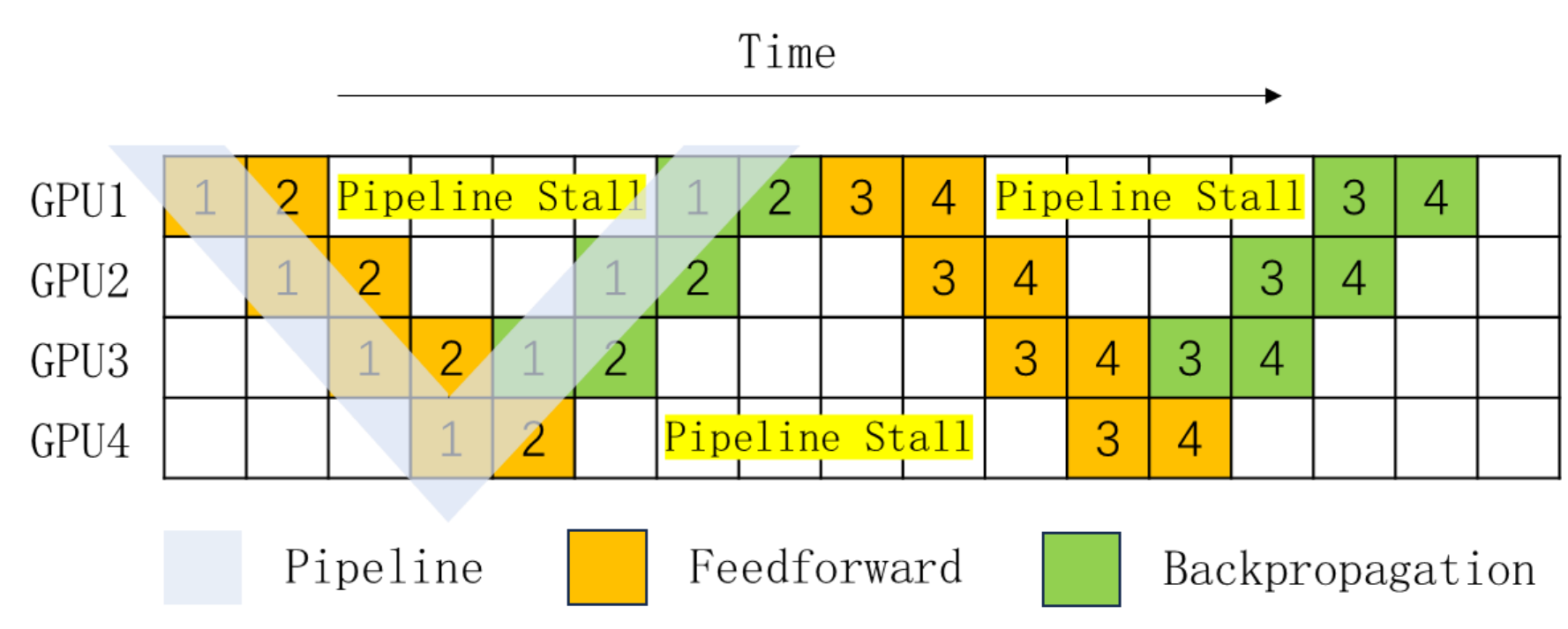}%
\label{fig:parallelism_pipeline}}
\caption{Parallelism modes of distributed DL}
\label{fig:parellelism}
\end{figure*}

Distributed DL partitions data and models into multiple processing units (typically GPUs) for parallel execution to leverage the computational capacity of many computing nodes in a cluster. 
As illustrated in Fig.~\ref{fig:parellelism}, distributed DL has three basic parallelism modes: data parallelism, model parallelism, and pipeline parallelism.


\subsubsection{Data parallelism} 
As illustrated in Fig.~\ref{fig:parallelism_data}, data-parallel training partitions the entire training dataset into several splits and distributes them across many GPUs for parallel training~\cite{dataParallelPytorch20}.
Each GPU has a replicate of a whole model with an identical structure and trains it on a specific dataset partition. 
Throughout the distributed training process, these local models share their knowledge to update a global model using a specific model synchronization mechanism, usually via a parameter server (PS).

\subsubsection{Model parallelism} 
As illustrated in Fig.~\ref{fig:parallelism_model}, model-parallel training divides an entire DL model into submodels and distributes them onto many GPUs within a cluster when the model exceeds the capacity of a single GPU or computing node~\cite{modelParallelLargeModel22}.
This parallelism mode concerns the model division strategy focusing on workload balance and the submodel placement strategy focusing on communication overhead between submodels. 

\subsubsection{Pipeline parallelism} 
As illustrated in Fig.~\ref{fig:parallelism_pipeline}, pipeline-parallel training enhances DL parallelism by ordering different stages of distributed training in a pipeline and preventing computational and communication resources from idling~\cite{li2021chimera,liu2022autopipe,oh2022out}.
Pipeline-parallel training can be considered a special case of model-parallel training by decomposing the training of submodels layer by layer into subtasks and overlapping their computation of different stages across different GPUs. 
Computational and communication tasks can also overlap in the pipeline. 
This mode is typically applicable in the domains of LLM training~\cite{llmGpuCluster21}, edge computing~\cite{edgeCloudAI22}, and the Internet of Things~\cite{aiIoTSurvey22}, where devices have heterogeneous computational and communication capabilities to handle various distributed DL subtasks.   
In practice, pipeline parallelism can work with other parallelism modes to tackle complex distributed DL workloads with large model structures~\cite{tarnawski2021piper,swarmParallel23}.

\begin{figure*}[!t]
\centering
\includegraphics[width=1\textwidth]{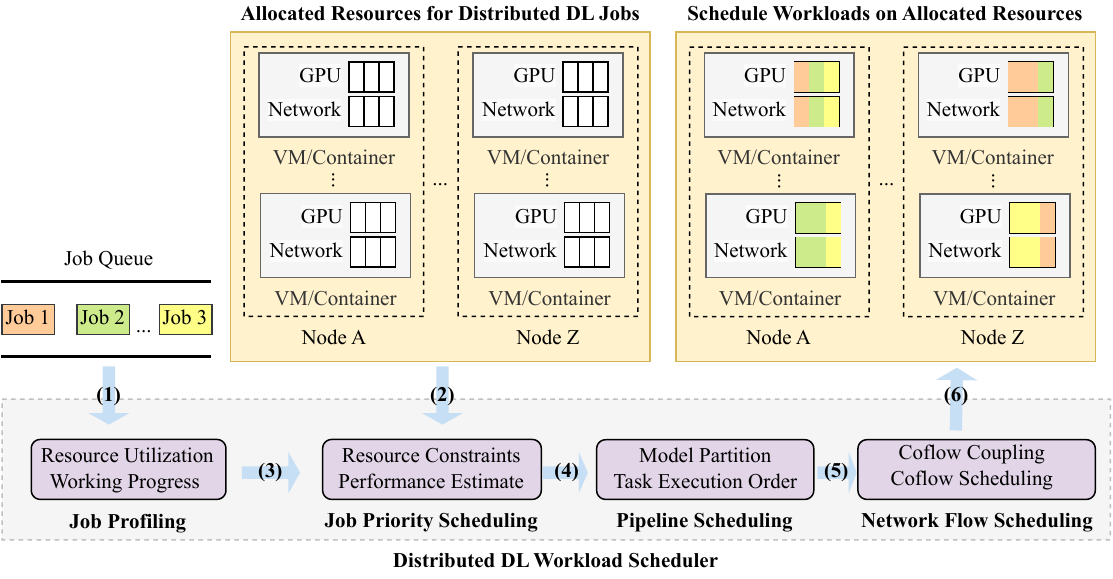}
\caption{An overview illustration of resource allocation and workload scheduling mechanisms in distributed DL}
\label{fig:resource_task}
\end{figure*}

\subsection{Resource Allocation and Workload Scheduling for Distributed DL}\label{sec:resourceTask}
Resource allocation and workload scheduling strategies for high-performance distributed DL are typically integrated in cluster-level and distributed-DL-level frameworks.
Fig.~\ref{fig:resource_task} illustrates the procedure of resource allocation and workload scheduling mechanisms for large-scale distributed DL within a cluster. 
This procedure comprises six major steps.
(1) For a queue of distributed DL jobs, the task scheduler conducts job profiling based on various workload characters, such as resource-utilization status and working progress. 
(2) The resource manager allocates GPU and network resources distributed within the cluster for jobs based on their characteristic profiling. The resources can be represented in a physical or virtual manner, and virtual resources can be encapsulated in virtual machines or containers.
(3) The job-level scheduler determines job-execution priorities based on resource constraints and job performance estimation.
(4) The pipeline-level scheduler divides the job into subtasks and locates them onto available resources for the pipeline execution of the subtasks, aiming to increase task parallelism and overlap computational and communication workloads.
(5) The network-flow-level scheduler optimizes the network flows or coflows of numerous subtasks by considering the relation and dependency of network flows.
(6) Scheduled jobs, pipelines, and network flows run efficiently on the allocated GPU and network resources within the cluster.


\newcommand\cb[1]{\colorbox{green!15}{\hyperref[#1]{[#1]}}}
\setlength\fboxsep{0pt}

\newcommand\gCate[1]{\parbox[t]{1.7cm}{\centering #1}}
\newcommand\gRef[1]{\parbox[t]{1.1cm}{#1}}
\newcommand\gYear[1]{\parbox[t]{0.4cm}{\centering #1}}
\newcommand\gContent[1]{\parbox[t]{8.5cm}{ #1 \vspace{1.5pt}}}
\begin{table*}[!t]

\tiny
        \caption{Studies on Resource Allocation Strategies for Large-Scale Distributed DL}\label{tab:GPUmanage}
        \centering
        \begin{tabular}{|c|c||l|c|l|}
                \Xhline{2\arrayrulewidth}
                \multicolumn{2}{|c||}{\cellcolor{green!15}\textbf{Category}} & \cellcolor{green!15}\textbf{Ref.} & \cellcolor{green!15}\textbf{Year} & \cellcolor{green!15}\textbf{Highlight} \\ 
                \Xhline{2\arrayrulewidth}
                \multirow{30}{*}{\rotatebox[origin=c]{90}{\parbox[t]{3cm}{\centering Training (\ref{sec:resource_train})}}} & \multirow{2}{*}{\gCate{GPU Sharing. Approach: (1) Workload Profiling \cb{C1} (\ref{sec:gpu_train})}} & 
                
                \gRef{Gandiva~\cite{xiao2018gandiva}} & \gYear{2018} & \gContent{ Using the profiles of the DL workload to improve efficiency of training DL models and latency in a GPU cluster.}\\
                \cline{3-5} 
                ~ & ~ & \gRef{AntMan~\cite{xiao2020antman}} & \gYear{2020} & \gContent{ Introducing co-designing the cluster scheduler and dynamic scaling mechanisms.} \\ 
                \cline{3-5}
                ~ & ~ & \gRef{FGD \cite{weng2023beware}} & \gYear{2023} & \gContent{ Monitoring the individual evaluation functions of DL jobs at runtime to make placement decisions and resource allocations elastically.} \\  
                \cline{3-5}
                ~ &  & \gRef{TGS \cite{wu2023transparent}} & \gYear{2023} & \gContent{ Designing adaptive rate-control and transparent unified-memory mechanisms .} \\
                \cline{3-5}
                ~ & ~ & \gRef{Orion \cite{strati2024orion}} & \gYear{2024} & \gContent{Co-scheduling GPU kernels based on the computation and memory profiles of DNN workloads.} \\  
                \cline{2-5} 

                ~ & \multirow{2}{*}{\gCate{(2) Context Switching \cb{C2} (\ref{sec:gpu_train})}} & \gRef{Salus \cite{yu2019salus}} & \gYear{2019} & \gContent{ Achieving fine-grained GPU sharing among multiple DL applications.} \\ 
                \cline{3-5}
                ~ & ~ & \gRef{PipeSwitch \cite{bai2020pipeswitch}} & \gYear{2020} & \gContent{ Exploiting the profiles of DL applications to achieve millisecond-scale context switching.} \\ 
                \cline{3-5}
                ~ & ~ & \gRef{DistMind \cite{jin2024efficient}} & \gYear{2024} & \gContent{Exposing the abstractions of a GPU pool and a memory pool and designing high-performance three-stage pipelining.}    \\ 
                \cline{3-5}
                ~ & ~ & \gRef{G-Safe \cite{pavlidakis2024g}} & \gYear{2024} & \gContent{Offering transparent memory protection for context switching.} \\ 
                \cline{2-5}

                ~ & \multirow{2}{*}{\gCate{(3) Performance Estimating \cb{C3} (\ref{sec:gpu_train})}} & \gRef{Optimus~\cite{peng2018optimus}} & \gYear{2018} & \gContent{ Estimating a DL task’s remaining execution time and designing a marginal gain-based allocation algorithm.} \\
                \cline{3-5}
                ~ & ~ & \gRef{Harmony~\cite{Harmony2019}} & \gYear{2019} & \gContent{ Placing training jobs in a manner that minimizes interference  and maximizes performance.} \\  
                \cline{3-5}
                ~ & ~ & \gRef{Horus \cite{yeung2021horus}} & \gYear{2021} & \gContent{ Proposing a data-driven approach to predict the GPU utilization of heterogeneous DL tasks.} \\  
                \cline{3-5}
                ~ & ~ & \gRef{GPARS~\cite{wang2024gpars}} & \gYear{2024} & \gContent{Leveraging spatiotemporal correlations among jobs to allocate suitable GPU types for newly submitted jobs.} \\  

                \cline{2-5}
                ~ & \gCate{(4) Elastic Scaling \cb{C4} (\ref{sec:gpu_train})} & \gRef{Pollux \cite{qiao2021pollux}} & \gYear{2021} & \gContent{ Combining system throughput with statistical efficiency and introducing a formulation of goodput.} \\  
                \cline{3-5}
                ~ & ~ & \gRef{Zico \cite{2021Zico}} & \gYear{2021} & \gContent{ Monitoring the memory-usage pattern of individual DL jobs by tracking computational progress of training jobs.} \\  
                \cline{3-5}
                ~ & ~ & \gRef{AFS \cite{hwang2021elastic}} & \gYear{2021} & \gContent{ Handling future jobs requires proactive preparation based on current share calculations.} \\  
                \cline{3-5}
                \cline{3-5}
                ~ & ~ & \gRef{EasyScale~\cite{li2023easyscale}} & \gYear{2023} & \gContent{Utilizing a thread abstraction called EasyScaleThread to preserve consistent training accuracy with scalable GPUs.} \\ 
                ~ & ~ & \gRef{FlowCon~\cite{FlowCon2023}} & \gYear{2023} & \gContent{ Minimizing the growth of GPU fragmentation through packing tasks.} \\   
                \cline{2-5}

                ~ & \multirow{2}{*}{\gCate{(5) For Hyperparameter Tuning \cb{C5} (\ref{sec:gpu_train})}} & \gRef{Fluid \cite{MLSYS2021c0987e6b}} & \gYear{2021} & \gContent{ Utilizing a water-filling approach to accelerate the hyperparameter optimization process.} \\  
                \cline{3-5}
                ~ & ~ & \gRef{Titan \cite{gao2022titan}} & \gYear{2022} & \gContent{ Merging several fine-tuning workloads into one to improve resource utilization.} \\  
                \cline{3-5}
                ~ & ~ & \gRef{DISC \cite{liu2022adaptive}} & \gYear{2022} & \gContent{ Leveraging adaptive scaling to adjust the size of GPU time slices and formalizing the dynamic allocation of GPU time slices into an optimization problem.}\\
                \cline{3-5}
                ~ & ~ & \gRef{Hydro \cite{hu2023hydro}} & \gYear{2023} & \gContent{ Expanding resources for hyperparameter tuning workloads by interleaving them with pipeline-enabled large-model training tasks.} \\  
                \cline{2-5}

                ~ & \multirow{2}{*}{\gCate{Network Bandwidth Sharing. Granularity: (1) Job; (2) Gradient Block; (3) Coflow \cb{C6} (\ref{sec:network_allocate})}} &  
                \gRef{Liquid \cite{gu2021liquid}} & \gYear{2021} & \gContent{ (1) Proposing intelligent cluster network-efficient scheduling methods in both immediate and batch modes.} \\   
                \cline{3-5}
                ~ & ~ & \gRef{Prophet \cite{zhang2021prophet}} & \gYear{2021} & \gContent{ (2) Employing the monitored network bandwidth and the profiled gradient time interval to predict the number of gradients into gradient blocks.} \\  
                \cline{3-5}
                ~ & ~ & \gRef{Parrot \cite{li2020efficient}} & \gYear{2020} & \gContent{ (3) Using a linear program (LP) solution to derive a weighted bandwidth scaling
                        strategy to minimize the time cost in the communication stage.} \\ 
                
                \Xhline{2\arrayrulewidth}
                \multirow{11}{*}{\rotatebox[origin=c]{90}{\parbox[t]{2.5cm}{\centering Inference (\ref{sec:GPU_inference})}}} & \multirow{2}{*}{\gCate{Spatial Sharing \cb{C7}}} & \gRef{GSLICE \cite{dhakal2020gslice}} & \gYear{2020} & \gContent{ Developing self-learning and adaptive GPU-resource allocation and batching schemes.} \\ \cline{3-5}
                ~ &  & \gRef{iGniter \cite{xu2022igniter}} & \gYear{2022} & \gContent{ Leveraging inference performance model to calculate the appropriate batch size and lower bound of allocated GPU resources.} \\ \cline{3-5}
                ~ & ~ & \gRef{SLO-aware \cite{cho2022sla}} & \gYear{2022} & \gContent{ Distributing inference requests to the deployed functions based on the autoscaling decision.}\\
                \cline{3-5}
                ~ & ~ & \gRef{AlpaServe~\cite{li2023alpaserve}} & \gYear{2023} & \gContent{Finding a partitioning strategy that minimizes the stage imbalance for inter-operator parallelism.}  \\
                \cline{2-5}
                ~ & \multirow{2}{*}{\gCate{Temporal Sharing \cb{C7}} } & \gRef{Nexus \cite{shen2019nexus}} & \gYear{2019} & \gContent{Applying a heuristic approach to select the requests.} \\ \cline{3-5}
                ~ & ~ & \gRef{INFaaS \cite{romero2021infaas}} & \gYear{2021} & \gContent{ Identifying the colocation interference caused by the shared hardware resources.}\\
                \cline{3-5}
                ~ &  & \gRef{Cocktail \cite{gunasekaran2022cocktail}} & \gYear{2022} & \gContent{ Building a distributed-weighted auto-scaling policy that utilizes the importance sampling technique.}\\
                \cline{2-5}
                ~ & \multirow{2}{*}{\gCate{Hybrid Sharing \cb{C7}}} & \gRef{Gpulet \cite{choi2022serving}} & \gYear{2022} & \gContent{ Allowing heterogeneous ML models to be mapped to multiple gpulets in the most cost-effective way.} \\ \cline{3-5}
                ~ &  & \gRef{FaST-GShare~\cite{gu2023fast}} & \gYear{2023} & \gContent{ Introducing the FaST-Manager to limit and isolate spatio-temporal resources.}\\
                \Xhline{2\arrayrulewidth}
        \end{tabular}
\end{table*}

\section{Resource Allocation}\label{sec:resourceAlloc}
In this section, we introduce resource allocation strategies for both distributed training and inference, which have different workload characteristics and performance requirements. 
Table~\ref{tab:GPUmanage} summarizes these strategies with related challenges \cb{Cx}, highlighted in box texts in the coming sections.
Resource allocation strategies for distributed training are classified into two categories: GPU sharing and network bandwidth sharing. 
GPU sharing strategies are further classified into five approaches based on techniques they applied, including workload profiling, context switching, performance estimating, elastic scaling, and special considerations for hyperparameter tuning workloads.
Network bandwidth sharing strategies are further classified based on the targeting granularity of resource, including the job, gradient block task, and coflow.  
Resource allocation strategies for distributed inference are classified into three categories based on sharing patterns of GPUs, including spatial, temporal, and hybrid sharing. 

\subsection{Resource Allocation for Distributed Training}\label{sec:resource_train}
The training process of distributed DL requires an intensive consumption of computational power and memory of GPUs and network communication bandwidth across GPUs. Therefore, GPU and network bandwidth sharing is the focus of the discussion on resource allocation for distributed training.

\subsubsection{GPU sharing:}\label{sec:gpu_train}
Although GPUs have found extensive applications in distributed DL, a prevalent issue of underutilization is observed in production clusters. The recorded GPU utilization typically ranges from 25\% to below 50\%~\cite{narayanan2020heterogeneity, hu2021characterization,weng2022mlaas,li2023lyra, weng2023beware, cheng2023towards}. This concern is particularly noteworthy in large-scale distributed computing environments. To address this issue, various distributed technologies have been developed to enable DL tasks to run efficiently on numerous devices~\cite{gao2022deep}. 

GPU sharing strategies typically leverage partial resource allocation through virtualization to mitigate the challenge of low GPU utilization in large-scale distributed DL. NVIDIA, acknowledged as the leading GPU provider, introduces Multiple Process Sharing (MPS)~\cite{NvidiaMPS} that offers an operating-system-level virtualization solution. Nevertheless, its implementation requires application-specific expertise to define resource limits for ensuring performance isolation. Moreover, MPS lacks compatibility with various DL frameworks. 
To address the performance isolation issue with MPS, another NVIDIA technology, Multi-Instance GPU (MIG)~\cite{NvidiaMIG}, enables the partitioning of a GPU into multiple discrete instances, each with dedicated resources. 
However, as MIG cannot dynamically adjust the partitions for GPU sharing to fit the GPU requirement of online workloads, it must initially allocate peak GPU resources for online workloads initially and retain them during the entire execution life cycle, leading to a significant waste of GPU resources.
To address the problems with efficient performance isolation for online workloads in MPS and MIG, Muxflow~\cite{zhao2023muxflow} proposes a two-level protection mechanism to guarantee GPU isolation for online workloads. 
The workload level protection resides between the CUDA~\cite{cuda} drive layer and CUDA runtime layer and controls the offline workloads to protect online workloads. 
The GPU level protection monitors the GPU device status to enable dynamic adjustment of the GPU memory quota for offline workloads. 
ByteDance has successfully deployed Muxflow in clusters with more than 20,000 GPUs.

\newcommand\challenge[1]{
\begin{table*}[htbp]
        \centering
        \begin{tabular}{|l|}
                \Xhline{1\arrayrulewidth}
                \parbox[t]{.973\textwidth}{
                \vspace{1pt}
                \cellcolor{green!5} #1 
                \vspace{6pt}
                }\\
                \cline{1-1}
        \end{tabular}
\end{table*}
}

\challenge{\textbf{Challenge [C1]:} Utilizing representative distributed DL workloads and profiling general characteristics so that the profiling result accurately reflects the workload characteristics of the working environment for the GPU-allocation strategy.}\label{C1}

$\bullet$ \textbf{Workload profiling:} Some solutions leverage the profiling of complex distributed DL workloads of production large-scale clusters or clouds to instruct the GPU allocation strategies, tackling Challenge \cb{C1}. 
Gandiva~\cite{xiao2018gandiva} leverages the profiles of distributed DL tasks and addresses the issue of GPU underutilization in three key ways. Initially, Gandiva allows incoming jobs to time-share GPUs with existing jobs when overloaded. Then, it permits time-sliced jobs to migrate to other GPUs. Lastly, it supports elastic GPU capacity, increasing the number of GPUs during idle times and reducing the number of GPUs as the load grows dynamically, thereby utilizing idle GPUs effectively. 
The performance of Gandiva is demonstrated on production clusters at Microsoft. 
AntMan~\cite{xiao2020antman} is a production solution for distributed DL clusters at Alibaba.
It analyzes the cause of GPU underutilization in distributed DL clusters for production use in three aspects: hardware, cluster scheduling, and job behavior. 
Exploiting the profiles of fluctuating resource demands from distributed training jobs, AntMan co-designs the cluster scheduler and distributed DL framework with dynamic scaling mechanisms for GPU resources during job execution. This approach ensures jobs' service-level objectives (SLOs) in large-scale clusters while enhancing cluster utilization through opportunistic scheduling.
Leveraging the analysis of the production trace at Alibaba, Fragmentation Gradient Descent (FGD)~\cite{weng2023beware} addresses severe GPU fragmentation in large clusters.
FGD minimizes GPU fragmentation growth through task packing to achieve maximum GPU allocation rates. 
TGS~\cite{wu2023transparent} provides transparent GPU sharing at OS layer for distributed DL tasks in production clusters of containers.
TGS addresses challenges of the lack of application profiling knowledge and the potential oversubscription of GPU memory during the sharing of GPU resources.
It tackles the first challenge by monitoring and controlling the rate of sending GPU kernels to the GPU for each container adaptively, aiming to maximize the rate of opportunistic jobs while not affecting that of production jobs.
It tackles the second challenge by unifying GPU memory and host memory in a single address space via CUDA unified-memory allocation~\cite{cudaUnifiedMem} that enables both performance isolation and transparency of GPU memory allocation. Oversubscribed memory of opportunistic jobs is evicted to the host memory automatically, ensuring the performance of production jobs. 
Strati~\textit{et al.}~\cite{strati2024orion} suggest that DNN workloads have numerous data-dependent operators with distinct computation and memory requirements. 
These individual operators can saturate the GPU computation units or memory bandwidth but often leave other resources underutilized. 
To address the issue of imbalanced utilization of GPU and other resources, they propose a fine-grained and interference-aware GPU allocator, named Orion, to co-schedule GPU kernels based on the computation and memory profiles of DNN workloads to optimize overall resource utilization.

\challenge{\textbf{Challenge [C2]:} Reducing the latency of GPU context switching for distributed DL workloads, which include offloading and loading of models and data}\label{C2}

$\bullet$ \textbf{Context switching:} Some work utilizes fast context switching to reduce GPU latency, tackling Challenge \cb{C2}. 
GPU context switching refers to a GPU switching between processes when it is executing multiple training jobs or tasks in parallel or sequence. 
Salus~\cite{yu2019salus} achieves low switching latency via a fine-grained GPU sharing policy that exposes two GPU sharing primitives: fast job switching and memory sharing. The former enables rapid preemption and efficient time sharing for the currently active DL job on a GPU, whereas the latter packs smaller distributed DL tasks on the same device to ensure high memory utilization and prevent memory fragmentation. 
In contrast, PipeSwitch~\cite{bai2020pipeswitch} supports fast-context switching for pipelines of distributed DL jobs. PipeSwitch optimizes the context switching overhead through model-aware grouping for pipelines and proactive allocating of GPU memory.
The model-aware grouping of layers aims to minimize the overhead of transferring the model between CPUs and GPUs during context switching. 
The proactive allocation of GPU memory for standby workers before it should be active expedites the speed of context switching. To prevent job interference, PipeSwitch enforces process-level isolation, by initialing a new separate process for each active-worker task. 
To minimize the overhead of loading an application from the memory pool to a GPU, DistMind~\cite{jin2024efficient} exposes the abstractions of the GPU pool and memory pool and incorporates three-stage pipelining, cache-aware load balancing, and DNN-aware sharding in the GPU scheduler to achieve low application loading overhead and high GPU efficiency. 
G-Safe~\cite{pavlidakis2024g} focuses on the safety problem of GPU sharing in multi-tenant environments. It constrains the GPU kernels of each application to stay within the memory partition allocated to them during context switching.

\challenge{\textbf{Challenge [C3]:} Ensuring that the intermediately defined performance goal used for guiding the resource allocation strategy leads to a straightforward improvement in the actual performance goal when estimating the performance of distributed DL jobs in the GPU-allocation strategy.}\label{C3}

$\bullet$ \textbf{Performance estimating:} Some works employ the performance-estimate-guided approach to enhance GPU-resource allocation, tackling Challenge \cb{C3}. 
Both the performance goal and performance-estimation method can vary in these approaches. 
To illustrate Challenge \cb{C3}, when a strategy aims to reduce the average job completion time but uses GPU utilization as an intermediate performance estimate, it must ensure that a higher GPU utilization leads to a reduced average job completion time.
Optimus~\cite{peng2018optimus} introduces a dynamic allocation algorithm based on marginal gains, estimating the remaining execution time of a distributed DL task. In this greedy policy, a job with a larger marginal gain will be allocated a higher quota of GPU resources. 
Harmony~\cite{Harmony2019} uses a deep reinforcement learning (DRL) algorithm to place distributed DL jobs on GPU resources that lead to the minimum training or inference time. The learning rewards for unseen placements are guided by historical allocation samples. 
Horus~\cite{yeung2021horus} builds a model to predict GPU utilization of heterogeneous distributed DL tasks from computation graph features. It identifies GPU utilization as a general proxy metric for making optimal placement decisions.
GPARS~\cite{wang2024gpars} leverages spatiotemporal correlations among jobs and adopts graph attention networks for precise job duration prediction. It designs a dynamic objective function to allocate suitable GPU types for newly submitted jobs.

\challenge{\textbf{Challenge [C4]:} Determining the timing and quota for expanding resources in elastic distributed training, which requires monitoring runtime performance statistics.}\label{C4}

$\bullet$ \textbf{Elastic training:} Elastic training, which involves expanding and shrinking resource capacity dynamically, is an important strategy to improve resource utilization and save costs for distributed DL in the cloud environment. 
Many studies tackle Challenge \cb{C4} and focus on elastic GPU memory allocation. 
For example, Pollux~\cite{qiao2021pollux} adjusts GPU resources available to distributed DL jobs dynamically, aiming to maximize the overall training goodput within the cluster. 
To improve the efficiency of GPU memory sharing, Zico~\cite{2021Zico} monitors the memory-usage patterns of individual distributed DL jobs by tracking computational progress during training. Based on the monitoring statistics, Zico allocates and deallocates memory among concurrent jobs automatically, ensuring no exceeding of the memory budget. 
AFS~\cite{hwang2021elastic} points out that handling future jobs requires proactive preparation of resources based on current share calculations. When the GPU scheduler estimates that the GPU contention will be heavy in the future, it allocates more resources to long-lasting jobs; otherwise it allocates more resources to short jobs.
EasyScale~\cite{li2023easyscale} utilizes a thread abstraction called EasyScaleThread to preserve consistent training accuracy when the number of workers changes in data-parallel training and proposes intra-job and inter-job GPU schedulers to scale in or out GPUs for workers dynamically. 
The intra-job scheduler proposes online GPU allocation proposals to the inter-job scheduler to maximize  distributed training throughput of a specific distributed DL job, and the inter-job scheduler approves or declines proposals based on marginal speedup and workload balancing considerations.
EasyScale can improve GPU utilization in heterogeneous GPU clusters with such two-level elastic GPU scheduling.
In contrast, some studies focus on elastic container resources. For instance, FlowCon~\cite{FlowCon2023} introduces a container placement strategy based on growth efficiency and dynamic resource configuration for elastic allocation and withdrawal of resources during runtime.

\challenge{\textbf{Challenge [C5]:} Improving GPU utilization for batches of multiple hyperparameter tuning jobs, which have mostly homogeneous workloads among different jobs to improve overall training throughput.}\label{C5}

$\bullet$ \textbf{Hyperparameter tuning:} Hyperparameter tuning workloads represent a batch of distributed jobs with highly similar workload characteristics, which can thus be leveraged for resource allocation. 
Several studies explore strategies for improving GPU utilization
during hyperparameter tuning in distributed DL clusters, tackling Challenge \cb{C5}.
Fluid~\cite{MLSYS2021c0987e6b} is a distributed DL hyperparameter tuning execution engine that abstracts the hyperparameter tuning process as a sequence of trial groups. It employs a water-filling approach to expedite the hyperparameter tuning process to enhance GPU utilization. 
Titan~\cite{gao2022titan} adopts a heuristic approach by consolidating multiple fine-tuning workloads into one, which is particularly advantageous considering that multiple fine-tuning workloads often share the same model parameters. 
DISC~\cite{liu2022adaptive} leverages adaptive scaling to adjust the size of GPU time slices occupied by hyperparameter-tuning jobs at runtime. 
This dynamic allocation of GPU time slices for each hyperparameter tuning job is based on its potential to create a steep increase in the model accuracy.
Hydro~\cite{hu2023hydro} addresses cluster-wise resource utilization and tuning efficiency by incorporating a heterogeneity-aware allocation strategy. This method extends the resources of hyperparameter-tuning workloads by interleaving them with pipeline-enabled large-model training tasks. By effectively utilizing idle time intervals on each node caused by the gaps between the forward and backward processing of micro-batches, Hydro enhances overall resource utilization and tuning efficiency in large-scale distributed DL clusters.

\challenge{\textbf{Challenge [C6]:} Network bandwidth allocation for distributed DL must be based on sufficient knowledge of the workload and coordination of bandwidth resource across the application and network layers.}\label{C6}

\subsubsection{Network bandwidth sharing:}\label{sec:network_allocate} 
In large-scale distributed environments, where communication is often the performance bottleneck, network bandwidth is another significant factor determining the efficiency of distributed training. 
To tackle Challenge \cb{C6}, network layers, e.g., the transport layer, usually work collaboratively with the application layer, and the network bandwidth allocator be implemented on either layer based on the level of tasks required to allocate network bandwidth.

$\bullet$ \textbf{Job:} Some work focuses on network bandwidth sharing for multiple distributed training jobs. 
For instance, Liquid~\cite{gu2021liquid} proposes a computational and communication-resource-estimation algorithm and a network-efficient job-placement strategy for distributed training jobs. 
The resource-estimation algorithm models resource requirements of distributed training jobs, including GPU computing power, GPU memory, and network bandwidth requirements. 
The job-placement strategy assigns distributed training jobs to a cluster of computing nodes and containers, finding a best-fit job placement solution that satisfies the estimated computational and communication-resource requirements and exhibits less GPU fragmentation and network communication cost across containers.

$\bullet$ \textbf{Gradient block task:} 
A distributed training job can be break down into multiple training tasks of gradient blocks.
Some work focuses on network bandwidth sharing at the granularity of the gradient block task.
For instance, Prophet~\cite{zhang2021prophet} groups into certain gradient blocks based on the profiled time interval and models the distributed training time in terms of the network bandwidth and order of network transfers of gradient blocks. 
Based on this model, Prophet searches for an optimal order of network transfers of gradient blocks, aiming to minimize the distributed training time. 
This optimal order of gradient block transfers optimizes both the network bandwidth sharing among gradient blocks and the overlapping between network transfers and GPU computation.

$\bullet$ \textbf{Coflow:} A coflow is an abstraction several network flows related to a specific communication task, e.g., several or a fraction of gradient transfers, and is usually scheduled on the transport layer. 
Some work focuses on network bandwidth sharing for coflows.
For instance, Parrot~\cite{li2020efficient} perceives the communication pattern of a distributed training job as a series of dependent coflows and estimates the remaining processing time of distributed training jobs based on the amount of information carried per coflow.
Parrot allocates network bandwidth to active coflows of concurrent jobs within the cluster, so that the effective completion time of coflows of the job with a shorter remaining processing time has a higher priority to be minimized.


\challenge{\textbf{Challenge [C7]:} Satisfying SLOs, such as latency and throughput, for distributed inference jobs of various complexity within specific resource constraints.}\label{C7}

\subsection{Resource Allocation for Distributed Inference}\label{sec:GPU_inference}
In contrast to distributed training that caters to long-term offline workloads, distributed inference typically demands real-time execution with more stringent requirements on latency and accuracy. This difference in demands requires resource allocation solutions to address the distinct characteristics of inference workloads effectively. 
In the distributed inference process, GPU sharing is the focus of research, which primarily faces Challenge \cb{C7}.
To tackle this challenge, various resource allocation methods can be divided into three major categories: spatial, temporal, and hybrid sharing. 
In the context of multiple distributed DL jobs, 
the spatial sharing of GPUs involves the sharing of GPU space partitions while the temporal sharing involves the sharing of computation time slices of an entire GPU.
Hybrid approaches combine techniques from these two categories.

$\bullet$ \textbf{Spatial Sharing:}\label{sec:SGPUSharing} Many existing works exploit spatial sharing of GPUs to optimize the performance of distributed inference tasks. GSLICE~\cite{dhakal2020gslice} introduces an inference system that achieves safe and efficient GPU sharing through spatial GPU multiplexing systematically. It utilizes MPS~\cite{NvidiaMPS}, a GPU spatial-multiplexing framework with virtualization, to handle various inference requests. iGniter \cite{xu2022igniter} employs an inference performance model to calculate an appropriate batch size and the lower bound of allocated GPU resources. Subsequently, it allocates GPU resources for each inference workload by employing a greedy approach to identify the placement GPU devices that can achieve minimal performance interference. 
The SLO-aware ML Inference Framework~\cite{cho2022sla} designs a resource auto-scaling strategy in the cloud by leveraging rich and precise workload-specific metrics, with a special consideration of the heterogeneity in the GPU computational capability. This effective and elastic management of resources ensures meeting the SLO for diverse inference workloads in the cloud.
Tackling the problem that large models may not be deployed on a single GPU, AlpaServe~\cite{li2023alpaserve} utilizes queuing theory to mathematically verify the benefits of model parallelism and searches for a partitioning strategy that minimizes the stage imbalance for inter-operator model parallelism.

$\bullet$ \textbf{Temporal Sharing:}\label{sec:TGPUSharing} 
Recent temporal-sharing approaches designed for specific distributed inference systems have shown improvements in GPU utilization, especially in cloud environments shared by numerous tenants. Nexus~\cite{shen2019nexus} employs a heuristic approach to select requests for co-location on the same GPU. Initially, it determines the most suitable batch size to meet throughput and SLO requirements for the existing inference workloads. Subsequently, Nexus identifies all possible combinations within a GPU's duty cycle on a single GPU in a best-fit manner, maximizing utilization without violating latency requirements. Focusing on inference services in the cloud, INFaaS~\cite{romero2021infaas} addresses the problem of co-location interference arising from shared hardware resources. It allocates available resources to interfered instances through workload migration or virtual-machine-level scaling, aiming to reduce monetary costs through GPU sharing while meeting latency requirements via virtual-machine-level scaling. Cocktail~\cite{gunasekaran2022cocktail} scales the virtual machine resources for various inference models in the cloud automatically and proactively based on the predicted workload and popularity of these models. This approach enhances the efficiency of resource allocation in distributed DL inference systems with a specific set of supported inference models.

$\bullet$ \textbf{Hybrid Sharing:}\label{sec:HGPUSharing} Several works study the hybrid GPU sharing approaches, considering both spatial and temporal sharing. Gpulet~\cite{choi2022serving} supports spatial sharing of GPUs via the abstraction of virtual GPUs that are split partitions derived from physical GPUs. 
Given allocated virtual GPU resources, Gpulet supports temporal sharing by scheduling the batch sizes of inference jobs of multiple tenants, with a goal to guarantee the SLO. This hybrid design enables cost-effective cloud-resource allocation for the inference of numerous heterogeneous DL models. FaST-GShare~\cite{gu2023fast} utilizes spatial and temporal sharing of GPUs to maximize inference function throughput in the Function-as-a-Service serverless architecture for distributed DL.
It supports auto-scaling of inference resources in the cloud based on the profiling of function throughput and resource allocation, maximizing GPU utilization while ensuring the SLO.

\subsection{Discussion}

\noindent $\bullet$ \textit{Fine-grained and elastic GPU allocation strategies are critical for improving GPU utilization.} 
Coarse-grained or even exclusive-access GPU allocation for individual distributed DL jobs is common in small clusters but can lead to extremely low GPU utilization in the data center environment. 
Fine-grained GPU allocation strategies for diverse distributed training and inference workloads, which share GPU computational resources for multiple jobs and subtasks, are crucial for improving GPU utilization, reducing memory fragmentation, and ensuring performance isolation. 
However, as resource requirements can fluctuate during the long-term distributed training process, elastic GPU allocation strategies are also important for fully utilizing GPU resources while maintaining SLO from the cloud providers' perspective.
Both strategy approaches require the support of virtualization technologies and a deep understanding of the workload characteristics of distributed DL. 
Moreover, the latter approach also requires knowing the runtime workload performance, which can be achieved by performance monitoring and dynamic adaptation techniques.

$\bullet$ \textit{High-performance large-scale distributed DL requires the orchestration of efficient allocation of GPU and network resources.} 
The allocation of network resources can frequently be overlooked as a bottleneck for efficient resource utilization in distributed DL.
Many resource allocation strategies of distributed DL focus on addressing computation issues, such as low utilization, load imbalance, and long queuing delays. However, with the increase of the cluster scale, the complexity of GPU network connections increases exponentially, and lack of consideration to efficient network-resource allocation can result in significant low job-execution performance of large-scale distributed DL. 
Efficient network bandwidth allocation strategies can alleviate communication contention.
Fully utilizing resources of both GPU and network bandwidth leads to enhanced overall performance of distributed training and inference on a large scale.

$\bullet$ \textit{Heterogeneity in resources and workloads is a significant consideration for effective resource allocation strategies for distributed DL.}
Heterogeneous resources and workloads are pervasive in the data center environment, which has large-scale resources and numerous tenants. 
On the one hand, Computing nodes and networks in various specifications and configurations introduce resource heterogeneity, and the heterogeneity that affects the performance of distributed DL most is in the heterogeneity in the GPU computation capacity and network protocol, bandwidth, and topology. 
Lack of consideration for resource heterogeneity can either underestimate the resource capacity of come resources and cause resource underutilization or overestimate the resource capacity of some other resources and cause resource contention. 
On the other hand, distributed DL workload characteristics of different tenants over different processing stages of the job in different periods can also be heterogeneous.
Lack of consideration for heterogeneous workloads can cause inaccurate estimation of workload performance, which results in inferior resource allocation decisions. 


\newcommand\tCate[1]{\parbox[t]{1.3cm}{\centering #1}}
\newcommand\tRef[1]{\parbox[t]{1.5cm}{#1}}
\newcommand\tYear[1]{\parbox[t]{0.4cm}{\centering #1}}
\newcommand\tContent[1]{\parbox[t]{8.6cm}{ #1 \vspace{1.5pt}}}
\begin{table}[!t]
\setlength\fboxsep{0pt}
\tiny
    \caption{Studies on Workload Scheduling Strategies for Large-Scale Distributed Training}\label{tab:scheduling}
    \centering
    \begin{tabular}{|c|c||l|c|l|}
    \Xhline{2\arrayrulewidth}
        \multicolumn{2}{|c||}{\cellcolor{green!15}\textbf{Category}} & \cellcolor{green!15}\textbf{Ref.} & \cellcolor{green!15}\tYear{\textbf{Year}} & \cellcolor{green!15}\textbf{Highlight} \\ 
        \Xhline{2\arrayrulewidth}
        \multirow{50}{*}{\rotatebox[origin=c]{90}{\parbox[t]{5.0cm}{\centering Training Scheduling (\ref{sec:TraningSch})}}} & \multirow{2}{*}{\tCate{Throughput: Job Level \cb{C8} (\ref{sec:TrainThroughput})}} & \tRef{Tiresias~\cite{Gu_Chowdhury_Shin_Zhu_Jeon_Qian_Liu_Guo_2019} \cb{C8(1)}} & \tYear{2019} & \tContent{Using LAS algorithm to prioritize jobs based on their service, a metric defined as the multiplication of requested GPU resources and execution time.} \\ 
        \cline{3-5}
        ~ & & \tRef{OSDL~\cite{WANG2022109191}~\cb{C8(2)}} & \tYear{2022} & \tContent{Designing job-placement and scheduling algorithms in hybrid networks with OCS and EPS.} \\ 
        \cline{3-5}
        ~ & & \tRef{Heet~\cite{mo2024heet} \cb{C8(2)}} & \tYear{2024} & \tContent{Measuring scaling efficiency on heterogeneous nodes and uses a price function to balance scaling and scheduling efficiency.} \\ 
        \cline{3-5}
        ~ & & \tRef{FfDL~\cite{Jayaram_Muthusamy_Dube_Ishakian_Wang_Herta_Boag_Arroyo_Tantawi_Verma_etal._2019} \cb{C8(2)}} & \tYear{2019} & \tContent{Using the operating lessons from the industry practice to guide the balance dependability with scalability, elasticity, flexibility, and efficiency.} \\ 
        \cline{3-5}
        ~ & & \tRef{Philly~\cite{jeon2019analysis} \cb{C8(2)}} & \tYear{2019} & \tContent{Correlating scheduler logs with logs from individual jobs and conducting a thorough analysis about the impact of gang scheduling and locality constraints on the queuing delay and job runtime.} \\ 
        \cline{3-5}
        ~ & & \tRef{E-LAS~\cite{Sultana_Chen_Xu_Yuan_2020} \cb{C8(2)}} & \tYear{2020} & \tContent{Using real-time epoch progress rates specific to distributed training jobs, as well as services obtained from the temporal and spatial domains, to guide scheduling decisions} \\ 
        \cline{3-5}
        ~ & & \tRef{CASSINI~\cite{rajasekaran2024cassini} \cb{C8(2)}} & \tYear{2024} & \tContent{Introducing a circle geometric abstraction to model communication workload patterns and shifting the angles of the circles to interleave workloads of different jobs on the same network link.} \\ 
        \cline{3-5}
        ~ & & \tRef{Liu~\textit{et al.}~\cite{liu2024sampling} \cb{C8(2)}} & \tYear{2024} & \tContent{Proportionally assigning job workloads on heterogeneous clusters for load balancing and high throughput.} \\ 
        \cline{3-5}
        ~ & & \tRef{AutoSched~\cite{gao2024autosched} \cb{C8(2)}} & \tYear{2024} & \tContent{Generating simulated workload trace to search for the best framework configuration for existing distributed training schedulers.} \\ 
        \cline{3-5}
        ~ & & \tRef{SMD~\cite{Yu_Wu_Ji_Liu_2021}~\cb{C8(3)}} & \tYear{2021} & \tContent{Allowing multiple jobs to compete for the communication bandwidth.} \\ 
        \cline{3-5}
        ~ & & \tRef{Sched$^2$~\cite{luan2019sched2}~\cb{C8(3)}} & \tYear{2019} & \tContent{Using DRL to perform smart locality-aware scheduling of distributed training jobs.} \\ 
        \cline{3-5}
        ~ & & \tRef{MLFS~\cite{Wang_Liu_Shen_2020} \cb{C8(3)}} & \tYear{2020} & \tContent{Leveraging the data from the heuristic scheduling method for training a DRL model and making decisions on job scheduling using this trained DRL model automatically.} \\ 
        \cline{3-5} 
        ~ & & \tRef{Yang~\textit{et al.}~\cite{yang2023meta}} & \tYear{2023} & \tContent{Utilizing a trainable performance model to guide the exploration of DRL and adaptively scheduling \texttt{all-reduce} communication workloads.} \\
        \cline{2-5}
        ~ & \multirow{2}{*}{\tCate{Throughput: Pipeline Level \cb{C9} (\ref{sec:TrainThroughput})}} & \tRef{GPipe~\cite{huang2019gpipe}} & \tYear{2019} & \tContent{Distributing layer-wise model partitions across multiple GPUs and splitting mini-batches into micro-batches for pipelining execution.} \\ 
        \cline{3-5}        ~ &  & \tRef{PipeDream~\cite{narayanan2019pipedream}} & \tYear{2019} & \tContent{Using a heuristic model to determine the workload partitioned on each worker to balance workloads and minimize communication overheads under various model and hardware constraints.} \\
        \cline{3-5}
        ~ &  & \tRef{AutoPipe~\cite{liu2022autopipe}} & \tYear{2022} & \tContent{An heuristic-based adaptive method to achieve balanced model partitioning.} \\ 
        \cline{3-5}
        ~ &  & \tRef{HetPipe~\cite{park2020hetpipe}} & \tYear{2020} & \tContent{Integrating pipeline parallelism with data parallelism in heterogeneous GPU clusters.} \\
        \cline{3-5}
        ~ &  & \tRef{Piper~\cite{tarnawski2021piper}} & \tYear{2021} & \tContent{ A fine-grained pipeline workload partitioning scheme integrating various parallelism modes, including the data, layer-wise model, and tensor-wise model parallelism} \\ 
        \cline{3-5}
        ~ &  & \tRef{MG\_WFBP~\cite{shi2021mgwfbp}} & \tYear{2021} & \tContent{Separating the computation of backpropagation into subtasks bounded by merged-gradient layers and overlapping it with the communication of model synchronization in data-parallel training.} \\ 
        \cline{3-5}
        ~ &  & \tRef{DeAR~\cite{zhang2023dear}} & \tYear{2023} & \tContent{Decoupling the \texttt{all-reduce} primitive into two continuous operations, which enables overlapping communication tasks with feedforward tasks and reducing the communication overhead of model synchronization in data-parallel training.} \\ 
        \cline{3-5}
        ~ & & \tRef{ScheMoE~\cite{shi2024schemoe}} & \tYear{2024} & \tContent{Partitioning input tokens to smaller tensors in \texttt{all-to-all} communications to overlap communication and computation in MoE models training.} \\ 
        \cline{3-5}
        ~ &  & \tRef{Chimera~\cite{li2021chimera}} & \tYear{2021} & \tContent{Integrating bidirectional pipelines improve 1F1B stage execution order to reduce pipeline stall and memory overhead.} \\ 
        \cline{3-5}
        ~ &  & \tRef{OOO BackProp~\cite{oh2022out}} & \tYear{2022} & \tContent{Leveraging gradient computation dependencies to reorder stage executions in the pipeline, prioritizing critical gradient computations.} \\ 
        \cline{3-5}
        ~ &  & \tRef{Hanayo~\cite{liu2023hanayo}} & \tYear{2023} & \tContent{Running multiple waves of stages in a pipeline to reduce pipeline stall with low memory overhead.} \\ 
        \cline{3-5}
        ~ & & \tRef{MixPipe~\cite{zhang2023mixpipe}} & \tYear{2023} & \tContent{Bidirectional pipelines for synchronous data-parallel training, with a mixed scheduling of 1F1B and 2F1B to balance memory and pipeline stall.} \\ 
        \cline{3-5}
        ~ & & \tRef{AdaPipe~\cite{adapipeAsplos24}} & \tYear{2024} & \tContent{Adaptive recomputation for different stages in a pipeline to maximize saved recomputation cost within memory limits.} \\ 
        \cline{2-5}

        ~ & \multirow{2}{*}{\tCate{Throughput: Network Flow Level \cb{C10} (\ref{sec:TrainThroughput})}} & \tRef{JPAS~\cite{Zhou_He_Luo_Yu_Sun_2020}} & \tYear{2020} & \tContent{Using a simple greedy mechanism to order all DDL jobs periodically.} \\ 
        \cline{3-5}
        ~ & & \tRef{Geryon~\cite{wang2020geryon}} & \tYear{2020} & \tContent{Employing multiple flows with varying priorities to transfer parameters of different urgency levels.} \\
        \cline{3-5}
        ~ &  & \tRef{TensorExpress~\cite{kang2020tensorexpress}} & \tYear{2020} & \tContent{Enables each switch to transmit tensor packets according to their priority using multiple queues.} \\ 
        \cline{3-5}
        ~ &  & \tRef{Beamer~\cite{he2021beamer}} & \tYear{2021} & \tContent{Reducing the SCT by considering stage information in its scheduling approach.} \\ 
        \cline{3-5}
        ~ &  & \tRef{Tereis~\cite{chen2023tereis}} & \tYear{2023} & \tContent{Exploring the utilization of idle GPU computational resources during data transmission periods.} \\ 
        \cline{3-5}
        ~ &  & \tRef{Mercury~\cite{duan2023accelerating}} & \tYear{2023} & \tContent{Working on data packet to shift priority scheduling to the transport layer.} \\ 
        \cline{2-5}

        ~ & \multirow{2}{*}{\tCate{Cost Efficiency \cb{C11} (\ref{sec:TrainSchCost})}}& \tRef{Cynthia~\cite{Zheng_Xu_Chen_Zhou_Liu_2019}} & \tYear{2019} & \tContent{Providing predictable distributed training performance and reducing the training budget.} \\ 
        \cline{3-5}
        ~ & & \tRef{FC$^2$~\cite{Ta_2019}} & \tYear{2019} & \tContent{A scheduler that recommends cost-effective cloud-resource allocations for distributed training tasks with a PS.} \\ 
        \cline{3-5}
        ~ & & \tRef{Jahani~\cite{Jahani_Lattuada_Ciavotta_Ardagna_Amaldi_Zhang_2019}} & \tYear{2019} & \tContent{Modeling the scheduling process as a MILP problem to reduce the leasing cost in a global manner while maintaining the job latency.} \\ 
        \cline{3-5}
        ~ & & \tRef{GPOEO~\cite{Wang_Zhang_Lai_Hao_Wang_2022}} & \tYear{2022} & \tContent{Saving power in GPU data centers and using a customized scheduler to orchestrate jobs.} \\ 
        \cline{3-5}
        ~ & & \tRef{STS~\cite{10328678}} & \tYear{2023} & \tContent{Exploiting the probability distribution of early termination and adapting the resource assignment during the execution of the jobs to minimize the expected energy cost} \\ 
        \cline{2-5}
        ~ & \multirow{2}{*}{\tCate{Deadline Guarantee \cb{C12} (\ref{sec:TrainSchDea})}} & \tRef{GENIE~\cite{8778770}} & \tYear{2020} & \tContent{Proposing a prediction model derived from lightweight profiling to estimate the processing rate and response latency for diverse DL workloads.} \\
        \cline{3-5}
        ~ &  & \tRef{Chronus~\cite{Gao_Ye_Sun_Wen_Zhang_2021}} & \tYear{2021} & \tContent{Providing deadline guarantee for SLO jobs and maximizing the performance of best-effort jobs.} \\ 
        \cline{3-5}
        ~ &  & \tRef{Hydra~\cite{10036352}} & \tYear{2023} & \tContent{Adopting a sampling approach that exploits the inherent iterative periodicity of DL jobs to estimate job completion times accurately on heterogeneous GPUs.} \\ 
        \cline{3-5}
        ~ &  & \tRef{UniSched~\cite{gao2024u}} & \tYear{2024} & \tContent{Jointly optimize job profiling, job scheduling, and resource allocation to guarantee deadline.} \\
        \Xhline{2\arrayrulewidth}
    \end{tabular}
\end{table}

\section{Workload Scheduling}\label{sec:taskSchedule}
In large-scale GPU clusters with complex network connections, scheduling distributed DL workloads effectively is critical for ensuring the high performance of task execution, optimal hardware utilization, and achievement of various scheduling objectives. Training and inference stages of distributed DL are widely recognized as particularly computation and communication-intensive~\cite{gao2022deep}. 
The following section studies workload scheduling strategies on training and inference workloads and focuses on providing efficient communication or overlapping computational and communication tasks for overall efficiency in large-scale distributed DL.

\subsection{Distributed Training Scheduling}\label{sec:TraningSch}
Efficient workload scheduling strategies are crucial for distributed training workloads, especially in large-scale settings with large data, models, and device scales. 
Large-scale distributed training involves iteratively executing massive computational tasks for feedforward and backpropagation calculation and communication tasks for data flowing and model synchronization.
It is a long-term and computation-intensive process that requires efficient scheduling strategies to improve execution parallelism and completion time and meet various performance goals. 
In this subsection,
we survey workload scheduling strategies of large-scale distributed training with various performance goals and scheduling granularity levels. 
Table~\ref{tab:scheduling} summarizes these strategies, which are categorized by various performance goals, including throughput, cost efficiency, and deadline guarantee goals, 
while the strategies focusing on distributed training throughput are further classified into three categories based on the scheduling granularity: the job, pipeline, and network flow.

\subsubsection{Throughput}\label{sec:TrainThroughput}
The throughput of distributed DL refers to the speed at which jobs or tasks are completed or the amount of work accomplished per unit of time.
It is one of the most critical performance goals of distributed training scheduling~\cite{gao2022deep} and is determined synthetically by various factors, including resource utilization, parallelism level, and communication overhead. 
Workload scheduling strategies usually work on the job, pipeline, and network flow levels to achieve high throughput.

\challenge{\textbf{Challenge [C8]:} (1) Online scheduling of distributed DL jobs whose arrival and completion times are unpredictable to achieve high throughput; (2) Resource-aware and workload-aware scheduling of distributed DL jobs in complicated, heterogeneous, or opaque resource structures; (3) Efficiently solving complex distributed DL workload scheduling problems with various resource constraints and workloads.}\label{C8}

$\bullet$ \textbf{Job-level scheduling.} Scheduling distributed training at the job level, which involves the reordering job execution priorities and the placement of jobs on GPUs, is one of the most common and effective scheduling approaches~\cite{Gu_Chowdhury_Shin_Zhu_Jeon_Qian_Liu_Guo_2019,Jayaram_Muthusamy_Dube_Ishakian_Wang_Herta_Boag_Arroyo_Tantawi_Verma_etal._2019}. 
Job-level scheduling for distributed training workloads faces several challenges as stated in Challenge \cb{C8}, which include several aspects: online scheduling, resource-aware scheduling, and complexity. 
For Challenge \cb{C8(1)} about online scheduling, 
on the one hand, the unpredictable job arrival time requires a prompt online scheduling decision for each job upon its arrival, which may trigger significant preemption overhead if the system allows preemptive scheduling. 
On the other hand, the complex workload characteristics and resource topology make it hard to predict the job completion time accurately, and an inaccurate estimate can impede the scheduling algorithm from achieving high throughput.
For Challenge \cb{C8(2)} about resource-aware scheduling, the distributed DL workload scheduler should match workloads with large-scale resources, especially when the network topology of GPUs and nodes is complicated, heterogeneous, and sometimes even opaque and unobservable, e.g., in a multi-available-zone cloud environment. 
For Challenge \cb{C8(3)} about scheduling complexity, the complexity of the job-placement problem can increase exponentially with the scale of the cluster with various resource constraints and workloads, requiring efficient and practical algorithms to find the optimal scheduling solution. 

Some studies refine the job priority algorithm to tackle the preemption problem of online job scheduling.
For example, Tiresias~\cite{Gu_Chowdhury_Shin_Zhu_Jeon_Qian_Liu_Guo_2019} draws inspiration from the classic Multi-Level Feedback Queue (MLFQ) algorithm~\cite{chowdhury2015efficient} and develops a priority discretization approach to mitigate issues related to frequent preemption. 
In addition, Tiresias uses a Least-Attained-Service (LAS) algorithm to prioritize jobs based on their service levels, which are quantified by the product of requested GPU resources and execution time, to avoid scheduling starvation.

Some studies utilize resource topology-aware and workload-aware scheduling algorithms to improve performance estimation. 
For resource topology-aware solutions, 
OSDL~\cite{WANG2022109191} designs algorithms for job placing and scheduling of distributed training jobs in hybrid networks with optical circuit switching (OCS) and electrical packet switching (EPS). 
The job placing algorithm utilizes the hybrid network topology information to use lightpaths reasonably, and the job scheduling algorithm jointly optimizes bandwidth requests of distributed training jobs in the OCS and EPS domains.
Heet~\cite{mo2024heet} proposes a 3D collaborative filtering method to accurately measure the scaling efficiency of all elastic configurations on heterogeneous nodes, substantially reducing profiling overhead. Meanwhile, Heet utilizes a price function to effectively balance scaling efficiency and scheduling efficiency.

For workload-aware solutions, 
FfDL~\cite{Jayaram_Muthusamy_Dube_Ishakian_Wang_Herta_Boag_Arroyo_Tantawi_Verma_etal._2019}, an open-source scheduling platform developed by IBM, incorporates operational insights from industry practices to strike a balance between dependability and scalability, while maintaining elasticity, flexibility, and efficiency. 
In a related study, Philly~\cite{jeon2019analysis} performs a comprehensive analysis by correlating logs of the scheduler with logs of individual jobs, examining the impact of gang scheduling and locality constraints on queuing delay and job completion time. Drawing on insights from this analysis, Philly advocates relaxing locality constraints to enhance job time efficiency.
Unlike the above methods, which rely on job completion time estimates or prior knowledge, E-LAS~\cite{Sultana_Chen_Xu_Yuan_2020} utilizes real-time epoch progress rates specific to distributed training jobs, combined with service metrics derived from temporal and spatial domains, to inform scheduling decisions. E-LAS surpasses Tiresias in training throughput by reducing the average completion time for distributed training jobs.
CASSINI~\cite{rajasekaran2024cassini} is a network-workload-aware distributed training job scheduler that uses a geometric circle abstraction with angular rotations to represent time shifts for communication workload patterns.
It schedules different time shifts to distribute communication workloads on network links, interleaves communication workloads on the same network link, and reduces job completion time.
Liu~\textit{et al.}~\cite{liu2024sampling} leverage proportional workload assignment on a heterogeneous GPU cluster to maximize distributed training throughput and minimize job completion time. 
To reduce the scheduling computational complexity, they propose constructing the sparsification of feasible solutions through sampling, which can significantly decrease the decision-making latency.

In addition to common workload schedulers, which schedule distributed training workloads, some studies explore various configurations of existing workload schedulers to find the best configuration for specific workloads.
For example, AutoSched~\cite{gao2024autosched} develops a workload generation engine to produce training workloads that can reveal future trace patterns, which facilitates accurate and efficient configuration tuning of distributed training workload schedulers. 
With the generated workload trace, AutoSched searches for the best configuration via a learnable causal model. AutoSched is supposed to be a general configuration-turning framework for various off-the-shelf distributed training schedulers, including Tiresias.

Several methods tackle the scheduling complexity by modeling the scheduling problem as an optimization problem and applying dynamic programming or DRL algorithms to solve the problem efficiently.
SMD~\cite{Yu_Wu_Ji_Liu_2021} presents a resource-scheduling analytical model that accommodates multiple jobs competing for communication bandwidth. This model treats the scheduling problem as a non-convex integer non-linear program with bin-packing constraints. 
SMD introduces an $\epsilon$-approximation algorithm for its resolution, termed the sum-of-ratios multi dimensional knapsack decomposition.
Sched$^2$~\cite{luan2019sched2} utilizes DRL to schedule distributed training jobs with a locality-aware approach. This method tries to understand both the locality sensitivity of jobs and the fragmentation condition of clusters comprehensively within the entire learning stack. Through this heightened awareness, the DRL model adjusts its scheduling decisions dynamically and adaptively, responding effectively to the varying locality sensitivities of individual jobs and the evolving state of cluster fragmentation.
MLFS~\cite{Wang_Liu_Shen_2020} employs data from heuristic scheduling methods to train a DRL model and subsequently uses this model to make informed decisions about job scheduling autonomously. 
Yang~\textit{et al.}~\cite{yang2023meta} propose a meta-learning-based DRL method to improve the job completion time by adaptively scheduling \texttt{all-reduce} communication workloads in data-parallel training.
To address the issue of massive samples in DRL and improve DRL efficiency, the proposed method trains a performance model to predict the training time and guide the DRL exploration strategy into an effective search space.

\challenge{\textbf{Challenge [C9]:} Partitioning workloads for load balancing across different workers and optimizing the execution order to reduce pipeline stall and memory overhead in pipeline-level scheduling for distributed training workloads.}\label{C9}

$\bullet$ \textbf{Pipeline-level scheduling.} 
In the pipeline parallelism mode of distributed training, pipeline-level scheduling divides training mini-batches into micro-batches and manages the sequential processing of micro-batch tasks within a pipeline architecture.
This level of scheduling is widely adopted by large-model distributed training jobs. 
This scheduling approach orchestrates computational and communication tasks across various stages in a pre-defined execution order and aims to improve their execution parallelism. 
As a result, the execution of computational and communication tasks of the same or different stages overlap, which increases the overall pipeline throughput.
This approach primarily faces Challenge \cb{C9}.
The pipeline stall refers to the phenomenon of a faster stage halting to wait for dependent slower stages to catch up, which can lead to low pipeline utilization and high memory overhead.
The memory overhead is the space required to retain the results of the feedforward phase in the memory for the later calculation of the backpropagation phase in each micro-batch.

Pipeline parallelism is the state-of-the-art approach for large-model training. 
GPipe~\cite{huang2019gpipe}, a pioneer in utilizing pipeline parallelism to train large models, distributes layer-wise model partitions across multiple GPUs and splits mini-batches into micro-batches for pipelining execution. 
It reduces the pipeline memory overhead by recomputing the activations of the feedforward phase again in the backpropagation phase.
This library can achieve nearly linear convergence speedups and offer the flexibility to scale to various DNN models of immense sizes efficiently.
However, GPipe assumes a partitioned model for pipelining is readily available or specified manually by users and does not design a model partitioning scheme for load balance. 

To design an efficient model partitioning scheme for pipeline-level scheduling, some studies focus on balancing workloads across workers with hardware constraints. 
PipeDream~\cite{narayanan2019pipedream} builds a heuristic model to determine the workload to be partitioned on each worker to balance workloads and minimize communication overheads.
The model considers various constraints, including the model scale, training iteration, device memory capacity, hardware topology and bandwidth, and number of workers, 
and the decision result relies on inputs from a short profiling run. 
To evenly distribute workload among worker, PipeDream also integrates data parallelism with pipeline parallelism at certain stages. 
AutoPipe~\cite{liu2022autopipe} introduces an adaptive method to achieve balanced partitioning. It first generates a relatively balanced model partition scheme through dynamic programming.
It then refines the scheme using the heuristic that pipeline completion time can be reduced by moving certain stages in the pipeline's critical execution path forward or backward in the timeline.
However, both PipeDream and AutoPipe focus only on the homogeneous GPU setting. 
To address pipeline load balancing in heterogeneous GPU clusters, 
HetPipe~\cite{park2020hetpipe} partitions large DNN models to minimize the maximum completion time of the partitions within heterogeneous GPU memory bounds of multiple virtual workers in the pipeline. To reduce the communication overhead of fully synchronous pipeline parallelism, HetPipe introduces a wave-synchronous-parallel approach to allow bounded model staleness within a wave of micro-batches but guarantees convergence.

Some studies focus on fine-grained workload pipelining schemes to maximize pipeline throughput. 
For instance, Piper~\cite{tarnawski2021piper} focuses on fine-grained model partitioning, while MG\_WFBP~\cite{shi2021mgwfbp}, DeAR~\cite{zhang2023dear} and ScheMoE~\cite{shi2024schemoe} focus on fine-grained overlapping of computational and communication tasks.
For fine-grained model partitioning, Piper~\cite{tarnawski2021piper} supports tensor-wise model parallelism in the model partitioning scheme for pipeline scheduling, which is not supported in prior work, in addition to data parallelism and layer-wise model parallelism.
It applies a two-level dynamic programming algorithm to search for the optimal partitioning of a DNN model to maximize pipeline throughput within memory constraints. 
With increased search space, Piper can find high-quality parallelism configurations with high pipeline throughput.
For fine-grained computation and communication overlapping, MG\_WFBP~\cite{shi2021mgwfbp} divides the calculation of a backpropagation task into numerous subtasks separated by merged-gradient layers, which stand as trigger points for model synchronization in data-parallel training. 
As a result, the communication of model synchronization in a subtask can overlap with the computation of backpropagation in a subsequent subtask.
In contrast, DeAR~\cite{zhang2023dear} decouples the \texttt{all-reduce} primitive into two continuous operations: \texttt{reduce-scatter} and \texttt{all-gather}. 
This decouple enables overlapping communication tasks of the previous stage with feedforward tasks of the next stage in the pipeline execution, reducing the communication overhead of model synchronization in data-parallel training. 
Focusing on the distributed training of mixture-of-experts models, which has the communication bottleneck caused by the \texttt{all-to-all} collective communication, ScheMoE~\cite{shi2024schemoe} pipelines \texttt{all-to-all} communications with expert computations by virtually partitioning input tokens to multiple smaller tensors to increase the chance of task overlapping.

Some studies focus on optimizing the pipeline execution order to reduce pipeline stall.
For example, Chimera~\cite{li2021chimera} applies bidirectional pipelines that are composed of two pipelines executing stages in reserve directions in a one-forward-one-backward (1F1B) manner.
Computational tasks of different micro-batches are mostly overlapped on different workers and the resultant bidirectional pipelines execute in a compacter manner than in PipeDream. 
Chimera also builds a model for determining the optimal number of pipeline stages and number of replicated pipelines, whose values rely on empirical results as inputs.
Out-Of-Order (OOO) BackProp~\cite{oh2022out} leverages gradient computation dependencies to reorder stage executions in the pipeline to maximize GPU-resource utilization. In data-parallel training, OOO reorders the sequence of gradient computations to maximize the overlap between computation and parameter communication. In pipeline-parallel training, it prioritizes critical gradient computations to minimize pipeline stall.

Some studies focus on reducing GPU memory consumption and recomputation cost for pipeline execution. 
On the one hand, though the bidirectional pipeline approach of Chimera can achieve low pipeline stall, it has multiple model replicas in two pipelines, which requires large GPU memory consumption. 
Hanayo~\cite{liu2023hanayo} mitigates the issue of excessive memory consumption by running multiple waves of forward and backward stages in a pipeline to reduce pipeline stall while not increasing GPU memory consumption. 
MixPipe~\cite{zhang2023mixpipe}, another bidirectional pipeline approach for synchronous data-parallel training, regulates a flexible number of micro-batches injected into the bidirectional pipelines to balance pipeline and device utilization. 
MixPipe also features a mixed scheduling of 1F1B and 2F1B to balance memory usage and pipeline stall.
On the other hand, though the recomputation strategy for the backward stage in the pipeline can relieve memory consumption, its cost can be non-negligible. 
To balance memory saving and computation cost in recomputation, AdaPipe~\cite{adapipeAsplos24} models the memory and time cost of different recomputation strategies and introduces an adaptive recomputation mechanism to allow different recomputation strategies, e.g., partial and full recomputation, for different stages in a pipeline. 
AdaPipe achieves maximum saved recomputation cost within memory limits.

\challenge{\textbf{Challenge [C10]:} Scheduling network flows at different granularity level to increase bandwidth utilization, network latency, and network congestion for distributed DL workloads.}\label{C10}
    
$\bullet$ \textbf{Network-flow-level scheduling.} 
Efficient network flow scheduling determines the transmission priority of data packets, network flows, and coflows related to distributed DL jobs, aiming to significantly increase network bandwidth utilization, reduce network latency, and avoid network congestion, as stated in Challenge \cb{C10}.
Network flow scheduling can work at various granularity levels, including the job, coflow, and data packet levels.

Some studies focus on the job level. 
JPAS~\cite{Zhou_He_Luo_Yu_Sun_2020} implements a straightforward greedy mechanism to organize all distributed training jobs periodically. This approach enables each host machine to prioritize its network flows according to the established job order, delegating the task of flow scheduling and rate allocation to the underlying priority-enabled networks.
Tereis~\cite{chen2023tereis} explores the utilization of idle GPU computational resources during data transmission periods. It predicts the completion time for a distributed DL job and its corresponding data transmission time, allowing for the simultaneous packaging of two jobs on the same GPU. This ensures that one job is completed before the other concludes its data transfer.

Some studies focus on the coflow level. 
Geryon~\cite{wang2020geryon} employs multiple flows with varying priorities to transfer parameters of different urgency levels. This approach coordinates multiple PSs effectively and gives precedence to urgent parameter transfers across the entire network fabric.
Beamer~\cite{he2021beamer} focuses on reducing the stage-completion time (SCT) by considering stage information in its scheduling approach. It proposes a stage-aware coflow-scheduling method to minimize the average SCT.

Some other studies focus on the data packet level. 
To address in-network delays, such as queuing delays, TensorExpress~\cite{kang2020tensorexpress} shifts priority scheduling to the transport layer, focusing on the packet granularity. 
It enables each switch to transmit tensor packets according to their priorities using multiple queues. This method ensures that high-priority data packets are handled efficiently to minimize delays.
Similarly, Mercury~\cite{duan2023accelerating} transmits packets with the highest priority in the Mercury buffer first. Additionally, Mercury incorporates immediate aggregation at the transport layer, enabling full overlapping of gradient push-and-pull operations. This approach not only streamlines data flow but also maximizes the efficiency of network resource utilization.

\challenge{\textbf{Challenge [C11]:} Jointly optimizing energy consumption, monetary cost, and throughput for distributed DL workloads with awareness of cloud resources and policies from the perspectives of cloud service providers or service users.}\label{C11}

\subsubsection{Cost efficiency}\label{sec:TrainSchCost}
The cost-efficiency objective of distributed training scheduling aims to minimize operational costs while ensuring optimal performance for distributed training workloads, especially in the cloud environment. 
It primarily faces Challenge \cb{C11} and focuses on a balance between resource utilization, energy consumption, and monetary expenditures in the scheduling decisions. 
Cynthia~\cite{Zheng_Xu_Chen_Zhou_Liu_2019} offers predictable distributed training performance while reducing the training budget. This scheduler identifies the optimal resource type and maintains training throughput effectively, thereby minimizing monetary costs.
Similar to Cynthia, FC$^2$~\cite{Ta_2019} is a scheduler that recommends cost-effective cloud resource allocations for parameter servers in distributed training tasks. It prioritizes instances with the largest network bandwidth within the budget to circumvent communication bottlenecks. Furthermore, it introduces a heuristic named Scale-Opt for determining worker instances, ensuring job throughput, and maximizing cost savings.
Jahani~\cite{Jahani_Lattuada_Ciavotta_Ardagna_Amaldi_Zhang_2019} considers computing nodes with varying numbers of GPUs as distinct virtual machines. The scheduling process is modeled as a mixed-integer linear programming (MILP) problem, aiming to reduce leasing costs globally while maintaining job latency.
GPOEO~\cite{Wang_Zhang_Lai_Hao_Wang_2022} achieves significant power savings for training workloads. It can be integrated into GPU data centers easily, utilizing a customized scheduler to manage job orchestration.
STS~\cite{10328678} optimizes the scheduling of distributed training jobs from the perspective of cloud service providers operating data centers. 
It leverages the probability distribution of early job termination to adapt resource assignments during job execution, with the aim of minimizing the expected energy cost.

\challenge{\textbf{Challenge [C12]:} Accurately estimating job completion or remaining times based on workload monitoring statistics in distributed DL workload scheduling to guarantee deadlines in the cloud environment.}\label{C12}

\subsubsection{Deadline Guarantee}\label{sec:TrainSchDea}
Deadline-guaranteed scheduling focuses on ensuring the completion of distributed DL jobs before a specified deadline for jobs whose timing is a crucial consideration. 
This performance goal is more common in the cloud environment, where cloud providers can elastically scale resources for distributed training workloads to guarantee the SLO for cloud users. 
Achieving this performance goal primarily faces Challenge \cb{C12}.
GENIE~\cite{8778770}, a trailblazing deadline-aware scheduler for distributed training workloads, explores the key factors that impact the performance of distributed DL tasks. It introduces a predictive model based on lightweight profiling, enabling an accurate estimation of the processing rate and response latency for a variety of distributed DL workloads. However, a significant limitation of GENIE is that it is unable to handle mixed workloads that include both deadline-sensitive tasks and best-effort tasks simultaneously~\cite{gao2022deep}.
Chronus~\cite{Gao_Ye_Sun_Wen_Zhang_2021}, an end-to-end scheduling system, meets SLOs by guaranteeing deadlines for SLO-aware jobs while also enhancing the performance of best-effort jobs. This dual-focused strategy enables Chronus to manage a wide range of workload requirements.
By extending these studies, Hydra~\cite{10036352} emerges as a dynamic and multifaceted scheduler to tackle various scheduling challenges, including adhering to deadlines and reducing job completion times. Hydra introduces an sampling approach leveraging the iterative periodicity inherent in distributed DL jobs. This technique enables precise estimation of job completion times in heterogeneous GPU environments, thereby improving efficiency and effectiveness of scheduling for various distributed DL workloads.
In contrast to other work that usually optimizes a specific scheduling stage to guarantee deadline for distributed training jobs, UniSched~\cite{gao2024u} adopts a mixed integer linear programming framework to jointly optimize job profiling, job scheduling, and resource allocation to satisfy various scheduling objectives, including the deadline SLO and latency. 
Two key components support the optimization of UniSched: an estimator for estimating job completion time and a selector for selecting jobs and allocating resources. 

\begin{table}[!t]
\setlength\fboxsep{0pt}
\tiny
    \caption{Studies on Workload Scheduling Strategies for Large-Scale Distributed Inference}\label{tab:scheduling_infer}
    \centering
    \begin{tabular}{|c|c||l|c|l|}
    \Xhline{2\arrayrulewidth}
        \multicolumn{2}{|c||}{\cellcolor{green!15}\textbf{Category}} & \cellcolor{green!15}\textbf{Ref.} & \cellcolor{green!15}\textbf{Year} & \cellcolor{green!15}\textbf{Highlight} \\ 
        \Xhline{2\arrayrulewidth}
        \multirow{12}{*}{\rotatebox[origin=c]{90}{\parbox[t]{3cm}{\centering Inference Scheduling (\ref{sec:InferenceSch})}}} & \multirow{2}{*}{\tCate{Latency and Cost Efficiency \cb{C13} (\ref{sec:InferenceSchEff})}} & \tRef{Sniper~\cite{10.1145/3489517.3530474}} & \tYear{2022} & \tContent{Using non-invasive performance characterization networks based on neural network similarity (NNS) to predict the inference time of DNNs accurately.} \\ \cline{3-5}
        ~ & ~ & \tRef{Ace-Sniper~\cite{10247268}} & \tYear{2024} & \tContent{Including both hardware and software platform information in the resource abstraction to tackle heterogeneous hardware and platforms for distributed inference.}\\
        \cline{3-5}
        ~ & & \tRef{AP$^{2}$~\cite{shi2023automatic}} & \tYear{2023} & \tContent{Minimizing distributed inference latency in 6G mobile communication systems with communications, heterogeneous devices, and task dependency constraints.} \\ 
        \cline{3-5}
        ~ & ~ & \tRef{AutoDeep~\cite{Li_Han_Zhang_Li_Tan_2020}} & \tYear{2020} & \tContent{Leveraging Bayesian Optimization and DRL to unearth the optimal cloud configuration and device placement with limited search time adaptively.}\\
        \cline{3-5}
        ~ & ~ & \tRef{HexGen~\cite{jiang2023hexgen}} & \tYear{2024} & \tContent{Applying asymmetric tensor-wise and layer-wise partitioning for pipeline-parallel inference to minimize communication and computation costs over heterogeneous GPUs.}\\
        \cline{2-5}

        ~ & \multirow{2}{*}{\tCate{Throughput \cb{C14} (\ref{sec:InferenceSchThr})}} & \tRef{Rafiki~\cite{Wang_Gao_Zhang_Wang_Chen_Ng_Ooi_Shao_Reyad_2018}} & \tYear{2018} & \tContent{Using a practical AIMD algorithm to adjust inference batch size.} \\ \cline{3-5}
        ~ & ~ & \tRef{Nanily~\cite{Tang_Wang_Liu_Wang_Han_2019}} & \tYear{2019} & \tContent{Deriving the corresponding batch size so that the inference completion time is equal to or close to the maximum remaining time.}\\
        \cline{3-5}
        ~ & ~ & \tRef{RRL~\cite{Qin_Zawad_Zhou_Yang_Zhao_Yan_2019}} & \tYear{2019} & \tContent{Focusing on optimizing parallel configurations at different levels.}\\
        \cline{3-5}
        ~ & ~ & \tRef{IRIS~\cite{ferikoglou2023iris}} & \tYear{2023} & \tContent{Adaptively adjusting the number of inference threads or containers based on predicted QPS to increase computation resource utilization in the cloud.}\\
        \cline{3-5}
        ~ & ~ & \tRef{Morphling~\cite{Wang_Yang_Yu_Wang_Li_Sun_He_Zhang_2021}} & \tYear{2021} & \tContent{Adapting the meta-model to a new inference service by sampling a small number of configurations and using it to find the optimal one.}\\
        \Xhline{2\arrayrulewidth}
    \end{tabular}
\end{table}

\subsection{Distributed Inference Scheduling}\label{sec:InferenceSch}
The scheduling of distributed inference workloads on available GPUs to meet various performance requirements is critical for the application of distributed DL models, especially as online services.
Distinct from distributed training workloads, which are typically iterative, long-term, and resource-intensive, distributed inference workloads exhibit another set of characteristics: one-round, short-term, and lightweight~\cite{gao2022deep, tang2023survey}.
In correspondence with such workload characteristic differences, the scheduling of distributed inference workloads also focuses on latency in addition to cost efficiency and throughput.
Table~\ref{tab:scheduling_infer} summarizes these distributed inference-scheduling strategies, focusing on various performance goals, including latency, cost efficiency, and throughput. 

\challenge{\textbf{Challenge [C13]:} Profiling workload characteristics of distributed inference jobs and scheduling them in low-latency and cost-efficient manners with an awareness of resource budgets.}\label{C13}

\subsubsection{Latency and cost efficiency}\label{sec:InferenceSchEff}
Scheduling distributed inference jobs faces Challenge \cb{C13}.
The inference latency refers to the time it takes to make a prediction given an inference query.
To maintain satisfactory latency, distributed inference schedulers are designed to scale resources proactively in response to request density and to reorder execution sequences strategically at the job level. 
For example,
Sniper~\cite{10.1145/3489517.3530474} stands out as a self-updating cloud-edge collaborative inference scheduling system with a focus on time awareness. It abstracts heterogeneous hardware resources and employs a non-invasive performance characterization model to predict the inference time of DNNs accurately based on neural network similarity. This system achieves a stable increase in throughput successfully even in dynamic cloud-edge environments, demonstrating its effectiveness and robustness in optimizing the distributed inference scheduling.
Ace-Sniper~\cite{10247268} extends Sniper by including software platform information in the resource abstraction, such as the CUDA and PyTorch library, to tackle heterogeneous hardware and platforms for distributed inference. 
Distributed inference latency is more of a concern in wireless networks, where communications are usually unstable and devices are heterogeneous. 
AP$^{2}$~\cite{shi2023automatic} aims to minimize distributed inference latency in 6G mobile communication systems with communications, heterogeneous devices, and task dependency constraints. 
It estimates task completion time on different devices based on profiling results and adopts a genetic algorithm~\cite{geneticAlgorithm16} to optimize the task arrangement for minimized inference latency while maintaining system reliability.

In practice, cost efficiency is another critical factor for distributed inference, especially when used in cloud services.
AutoDeep~\cite{Li_Han_Zhang_Li_Tan_2020} automates cloud deployment for real-time online DNN inference, focusing on minimizing costs while satisfying latency constraints. To achieve this, AutoDeep utilizes Bayesian optimization combined with DRL, which enables the adaptive discovery of the optimal cloud configuration and device placement and reduces the required searching time significantly. 
Through this method, AutoDeep achieves a trade-off between operational costs and latency in DNN inference workloads efficiently.
HexGen~\cite{jiang2023hexgen} improves distributed inference cost efficiency for large generative models over heterogeneous GPU devices. 
It applies asymmetric tensor-wise and layer-wise model partitioning for pipeline-parallel inference and aims to minimize communication and computation costs with heterogeneous GPU memory constraints. 

Latency and cost efficiency are recognized as interdependent objective in the inference system design. Improving one objective may inadvertently compromise the other if the solution is not designed meticulously, which motivates researchers to develop scheduling systems that optimizes both objectives simultaneously.
 
 \challenge{\textbf{Challenge [C14]:} Scheduling many distributed inference jobs with diverse workload characteristics to improve the inference throughput in the cloud.}\label{C14}

\subsubsection{Throughput}\label{sec:InferenceSchThr}
Scheduling batches of distributed inference jobs in the cloud also faces Challenge \cb{C14}.
To tackle this challenge, researchers typically refine the scheduling system for distributed inference workloads to enhance throughput through batch execution and configuration adjustments.

$\bullet$ \textbf{Batch execution:} Batching inference has been identified as an efficient method to enhance resource utilization and reduce scheduling overhead~\cite{gao2022deep}.  Various schedulers incorporate heuristic methods to fine-tune the batch size for the optimal performance.
For instance, Rafiki~\cite{Wang_Gao_Zhang_Wang_Chen_Ng_Ooi_Shao_Reyad_2018} employs a practical Additive-Increase Multiplicative-Decrease (AIMD) algorithm to adjust the inference batch size dynamically. This approach allows for responsive adaptation to varying workload conditions.
Nanily~\cite{Tang_Wang_Liu_Wang_Han_2019} establishes an upper limit on the batch size by calculating the maximum remaining time for a request, which is determined by subtracting the minimum queuing time of available resources from the remaining time. It then computes an appropriate batch size such that the inference completion time equals to or approximates this maximum remaining time.

$\bullet$ \textbf{Configuration adjustment:} In addition to the batch-execution approach, certain schedulers employ end-to-end configuration tuning to enhance distributed inference throughput.
RRL~\cite{Qin_Zawad_Zhou_Yang_Zhao_Yan_2019} emphasizes the optimization of parallel configurations at various levels, including inter-request-level and intra-request-level parallelisms. This optimization significantly reduces the overall system latency and improves the throughput.
In cloud environments, distributed inference throughput is significantly affected by client queries per second (QPS) and the number of parallel workers in the inference system. 
IRIS~\cite{ferikoglou2023iris} adaptively adjusts the parallelism level based on the online inference QPS predicted by a model pre-trained with monitoring profiles in an offline phase. 
IRIS integrates the parallelism level scheduling algorithm into the container orchestration platform, increasing overall computational resource utilization and throughput for distributed inference within the cluster.
Morphling~\cite{Wang_Yang_Yu_Wang_Li_Sun_He_Zhang_2021}, on the other hand, presents a rapid and near-optimal auto-configuration framework designed specifically for cloud-native model serving. This framework adapts to new inference services by sampling a limited set of configurations and then employs a meta-model to identify the most optimal configuration. This strategy allows Morphling to adjust quickly and efficiently to various service requirements while maintaining high system performance.

\subsection{Discussion}

$\bullet$ \textit{Fine-grained workload ordering and overlapping is key to scheduling large-scale distributed DL workloads in various parallelism modes.}
Job-level scheduling is important for online instant workloads and offline batch workloads when the distributed DL workloads are relatively lightweight in the multi-tenant data center environment. 
However, as the volumes of models and resources increase rapidly, pipeline-level scheduling for large models in large-scale clusters, a scheduling approach orthogonal to job-level scheduling, is essential for contemporary distributed training.
Though obeying different training procedures, data-parallel and model-parallel training can both leverage the pipeline to optimize the execution order and maximize the overlapping of different processing stages, including computation-computation, computation-communication, and communication-communication overlapping.
In practice, people pipelines the workloads of hybrid training parallelism modes for greater training throughput.

$\bullet$ \textit{Solving complex distributed DL workload scheduling problems typically requires DRL.}
Distributed DL workload scheduling can be a complex problem, especially in large-scale data centers with various resource and performance goal constraints.
Firstly, DRL can quickly adapt to constantly changing distributed DL environments and workloads by optimizing the policy through trial and error. 
Secondly, DRL can internally train DNNs for policy and value decisions to efficiently explore the vast search space of large-scale distributed DL workload scheduling.
Thirdly, DRL can make real-time decisions for distributed DL workload scheduling.

$\bullet$ \textit{Though throughput is essential for distributed DL workload scheduling, cost efficiency is an increasing concern.}
As the energy and monetary cost of distributed training and inference increases exponentially with large datasets and models, cost efficiency has become a decisive factor when deploying a training or inference process in the cloud from both providers' and users' perspectives. 
On the one hand, cloud providers must measure the cost of scheduling dynamic distributed DL workloads on diverse resources and design a competitive cost model for distributed training and inference services. 
On the other hand, cloud users need to estimate the cost of distributed training or inference based on the cost model and strike a balance between cost and other performance goals, such as throughput.


\section{Distributed Training of LLMs: A Case Study}\label{sec:caseStudy}
Recently, with the tremendous success of the application of LLMs~\cite{gpt20,lamda22,llama23,llama2_2023,paLM23} in various domains, such as NLP~\cite{llm2023survey}, programming~\cite{codeGen23}, finance~\cite{huang2023finbert}, and medicine~\cite{llmMedicine23NatureMedicine}, distributed training and fine-tuning LLMs efficiently have become a heated and important topic for researchers in the fields of computer science, artificial intelligence, and communications. 
As contemporary LLMs are in ultra-large sizes with up to hundreds of billions of parameters, training them typically requires hundreds of billions of tokens in the training dataset, hundreds of GPUs, and tens of days~\cite{llm2023survey}.
Efficient resource allocation and workload scheduling distributed DL strategies that can scale well to large data centers and workloads are critical for LLM training. 
This section examines several real cases of training LLMs in existing literature~\cite{llmGpuCluster21,ascendHpca21,decentralizedLlm22,lowCostdLlm22,llmMegatron22,swarmParallel23,OobleckResilientLlm23} to uncover insights and practical considerations for applying these distributed DL framework strategies in a large-scale setting.

 

$\bullet$ \textit{What are the important considerations for allocating resources across multiple data centers for LLM training?}
As LLM training requires a large volume of computational resources, the resources available in a single data center may not be able to support an LLM training job. 
Collaborative training of LLMs across multiple geologically apart data centers, , which form a \textit{computational power network} that share information and resources, has become a common practice~\cite{decentralizedLlm22} and faced several new challenges.
Firstly, compared to distributed training within a single data center, resource allocation across data centers faces significant challenges due to the heterogeneity in various computational and communication resources. 
Secondly, with many tenants across multiple data centers, performance isolation enabled by various virtual technologies must be ensured to prevent performance interference between different tenants and workloads. 
Thirdly, fault tolerance and data security are major concerns when considering the network transfer of the model and data to other data centers. The former concern will be discussed soon while the latter is solved mainly at the distributed DL algorithm level via federated learning~\cite{dmlToFL22}, which is not a focus of this survey.

$\bullet$ \textit{How to efficiently allocate resources for LLM training on a heterogeneous computational power network?}
Tackling heterogeneous resources in a computational power network, the resource allocator needs global knowledge about the resource capacity, pricing, and other specifications of all data centers~\cite{ascendHpca21}.
The computational power network should also monitor real-time resource usage status and workload profiles within each data center.
While leveraging both global and local resource information of the computational power network, the resource allocator considers user-specific and workload-specific requirements, such as the geological preference, price and completion time constraint, cost of transferring the training model and dataset, and training performance estimate. 
Based on these factors, the resource allocator can distribute black-boxed resources efficiently for LLM training within the computational power network with heterogeneous resources.

$\bullet$ \textit{How crucial is pipeline parallelism for LLM training?}
Pipeline parallelism is essential for LLM training.
As indicated in~\cite{llmGpuCluster21}, by diminishing the impact of the communication volume and worker idle time during pipeline flushes, heuristic pipeline parallelism proves effective in practice with trillion-scale LLMs on more than 3,000 GPU.
In contrast to layer-slicing parallelism, multiple-layer-slicing pipeline parallelism only communicates end-of-layer activations and gradients, which can be 300$times$ smaller in the communication volume in a 2.2-billion-parameter example~\cite{lowCostdLlm22}.
It is a common practice to use what is known as 3D parallelism~\cite{llmMegatron22,zeng2023distributed} for LLM training, which combines data, pipeline, and layer-slicing parallelisms, to maximize the training throughput. 


$\bullet$ \textit{How crucial is fault-tolerant scheduling for LLM training?}
Given the involvement of a large number of workers in prolonged training sessions for LLMs, ensuring fault-tolerance is of utmost importance for resilient scheduling. 
Frequent failure in devices or networks can potentially block the training process, degrade the convergence performance, and necessitate redundant restarting of failed tasks and pipelines. 
SWARM parallelism~\cite{swarmParallel23} incorporates the dynamic membership of unstable workers into account for fault-tolerant pipeline scheduling.
This dynamic fault-tolerant pipeline scheduling allows rerouting a task from a disconnected worker to other workers and ensures continuous task execution in case of worker failure in the pipeline.
According to Oobleck~\cite{OobleckResilientLlm23}, a failed pipeline can be recovered by using pipeline replicas and templates swiftly.
We can instantiate some logically equivalent pipeline replicas, which possess replicated model states. 
Additionally, we define pipeline templates, which include information about the number of workers and stages in the pipeline, as well as the mapping of stages to GPUs.
Once a pipeline failure occurs, a new pipeline can be restored based on the pipeline template and replicas instantly.


\section{Conclusion and Outlook}\label{sec:conclusion}
\subsection{Conclusion}
With the explosive increase in the volume of data, models, and resources, efficient framework strategies, including resource allocation and workload scheduling, are crucial for distributed DL.
This survey systematically investigates up-to-date efficient resource allocation and workload scheduling framework strategies for large-scale distributed DL. 
The discussion covers topics focusing on various resource types, scheduling granularity levels, and performance goals during the training and inference processes of distributed DL. 
We highlight the critical challenges for each topic and introduce the corresponding solutions.  
To illustrate the practical application of these framework strategies in real scenarios, we use a case study on distributed LLM training, typically with tens of billions of parameters on hundreds of GPUs.

\subsection{Outlook}
With the emergence of LLMs trained on ultra-large datasets, an increasing number of ultra-large GPU data centers are in use or on construction schedule.
Distributed DL framework strategies targeting large-scale settings with ultra-large data, models, and clusters are deemed a future research trend in this domain.
Compared to traditional distributed DL, large-scale distributed DL has new characteristics and poses new challenges for resource allocation and workload scheduling. 
We discuss these characteristics and challenges pertaining to the large scale as a hint of future research directions. 

\subsubsection{Multi-data center collaborative learning}
Many large technology corporations and research organizations are constructing computational power networks consisting of multiple geographically distributed GPU data centers. 
Promoting new computing paradigms for contemporary large-scale distributed DL, the computational power network enables the sharing and coordination of ultra-large computational resources across multiple data centers for large workloads.
However, compared to distributed DL within a single data center, the computational power network case presents various resource allocation and workload scheduling challenges, including higher resource heterogeneity, higher communication overhead, and tighter requirements for fault tolerance and data security. 
Overcoming these challenges requires scalable algorithms that work efficiently in large-scale environments with various constraints. 

\subsubsection{Resource and workload heterogeneity}
With the frequent upgrades of hardware devices for distributed DL, data centers commonly have heterogeneous computational and communication resources with various capacities in storage, computation speed, network bandwidth and latency, and energy and pricing costs.
Distributed DL workloads also exhibit heterogeneity in various aspects, such as dataset distribution, model complexity, distributed training parallelism modes, model synchronization mechanisms, and training dynamics.
A promising research direction is dynamic resource allocation to adjust resources based on resource availability across heterogeneous environments.
Another is adaptive workload scheduling, which matches the dynamic nature of changing workloads during prolonged training and inference processes to continually optimize performance.
Solving these scheduling problems with resource and workload heterogeneity also requires efficient optimization methods, such as DRL. 

\subsubsection{Pipeline execution for large-model training}
Pipelining has been applied to overlap various computational and communication workloads in large-model training. 
However, the optimization of pipeline execution for various training parallelism modes of large models has yet to be sufficiently explored. 
An example is related to adaptive pipeline scheduling. 
During the long-term execution of large-model training with dynamic workloads, the pipeline execution plan that initially balanced workloads can lead to significant workload skew among the workers. 
Fine-grained adaptive pipeline scheduling that dynamically adjusts the granularity of pipeline stages and rebalancing workloads can reduce pipeline stall throughout the training process. 
Another example is related to hierarchical pipeline scheduling.
To isolate the negative influence of the pipeline stall, the global and multi-level local workload schedulers can use multiple pipelines hierarchically within a computing node, within a rack, within a data center, and across data centers.

\subsubsection{Resilient distributed DL}
Failure in devices and tasks are always in company with distributed frameworks, and its influence is non-negligible in large-scale environments.
Resilient distributed DL framework strategies that can tolerate various failures become important for large-scale resource allocation and workload scheduling.
When optimizing distributed DL framework strategies, the solution should consider GPU failures, network disruptions, and storage issues that can interrupt the distributed DL process and decay performance.
Resource allocation strategies could proactively allocate redundant resources based on device failure considerations to tolerate failures.
Workload scheduling strategies could replicate datasets and tasks to tolerate task failures or conduct checkpoints to reduce recovery overhead. 

\subsubsection{Orchestration with distributed DL algorithms}
By being aware of the mechanisms of distributed DL algorithms, distributed DL framework strategies can be optimized for and orchestrated with them.
For example, the workload scheduler can accurately estimate the communication overhead and effectively optimize the scheduling solution by orchestrating lossless or lossy compression technologies for gradient compression and model synchronization mechanisms for data-parallel training and federating learning.
The resource allocator can promptly and dynamically adjust corresponding resources for distributed DL workloads by being aware of the adaptive policies of distributed DL algorithms.
Distributed DL algorithms can also be jointly optimized with distributed DL framework strategies for their best configurations and adaptive policies.

\subsubsection{Orchestration with distributed DL infrastructures}
Modern distributed DL infrastructures usually apply virtualization technologies and programmable network devices to extend resource capacities. 
On the one hand, virtualization technologies change resource capacities and pricing costs of computational and communication devices.
On the other hand, programmable network devices, such as programmable switches, extend the capability of network devices by integrating limited computational power.
They often work to optimize distributed DL-aware network traffic, such as in-network aggregation for distributed gradients. 
These infrastructure technologies allow framework strategies to be optimized for various performance goals other than throughput, such as cost efficiency, performance isolation, and network congestion.
Resource allocation strategies can leverage virtualization technologies to extend the allocation capability elastically, and workload scheduling strategies should consider virtualization performance and in-network aggregation when determining an optimal scheduling solution.


\bibliographystyle{ACM-Reference-Format}
\bibliography{sample-base,surveyBib,surveySyncBib,surveyOptBib,surveyGPUBib,surveySchedulingBib,surveyLlmBib}

\end{document}